\documentclass[12pt]{article}
\usepackage{graphicx}
\usepackage[centertags]{amsmath}
\usepackage{amssymb}
\usepackage{enumerate}
\usepackage{color}
\usepackage[sans]{dsfont}
\usepackage{colortbl}
\usepackage{paralist}
\usepackage{longtable}
\usepackage{ulem}
\usepackage{cancel}
\usepackage{color}
\usepackage{url}
\usepackage{mathrsfs}
\usepackage{multirow} %For tables
\usepackage{rotating} %Gives rotated table labels.
\usepackage{placeins}
\usepackage{subfigure}

\newcommand{\vev}[1]{\ensuremath{\langle #1 \rangle}}
\newcommand{\order}[1]{\ensuremath{\mathscr{O}\left(#1\right)}}

\newcommand{\SU}[1]{\ensuremath{\mathrm{SU}(#1)}}
\newcommand{\U}[1]{\ensuremath{\mathrm{U}(#1)}}

\newcommand{\slz}{\ensuremath{\mathrm{SL}(2,\mathbb{Z})}}

\newcommand{\gev}{{\rm ~GeV~}}

\newcommand{\mpl}{M_{Pl}}
\newcommand{\mstring}{M_{\textsc{s}}}
\newcommand{\msusy}{{M_{\textsc{susy}}}}
\newcommand{\mz}{M_{\textsc{z}}}

\newcommand{\re}[1]{\ensuremath{\text{Re} \ #1}}
\newcommand{\im}[1]{\ensuremath{\text{Im} \ #1}}

\newcommand{\lsqcd}{\ensuremath{\Lambda_{\textsc{sqcd}}}}

\newcommand{\fg}{\ensuremath{\mathcal{G}}}
\newcommand{\fk}{\ensuremath{\mathcal{K}}}

\newcommand{\fw}{\ensuremath{\mathcal{W}}}

\newcommand{\be}{\begin{equation}}
\newcommand{\ee}{\end{equation}}
\newcommand{\bea}{\begin{eqnarray}}
\newcommand{\eea}{\end{eqnarray}}

\newcommand{\zsix}{\ensuremath{\mathbb{Z}_6\text{-II }}}

\def\bctr{\begin{center}}
\def\ectr{\end{center}}

\begin{document}

\begin{titlepage}

\vspace*{-2cm}
\begin{flushright}
OHSTPY-HEP-T-09-003    \\
SU-ITP-10-05 \\
NSF-KITP-10-067
\end{flushright}
\vspace{1cm}

\begin{center}
{\bf Moduli stabilization and SUSY breaking \\ in heterotic orbifold
string models}

\vspace{2cm}

\texttt{ Ben Dundee\footnote[1]{Email:
\texttt{dundee@mps.ohio-state.edu}}{}$^{a\dagger}$, Stuart
Raby\footnote[2]{Email:
\texttt{raby@pacific.mps.ohio-state.edu}}{}$^{a\dagger}$, and Alexander
Westphal\footnote[3]{Email:
\texttt{awestpha@stanford.edu}}{}$^{b\dagger}$}
\\[5mm]
\textit{\small
{}$^a$ Department of Physics, The Ohio State University,\\
191 W. Woodruff Ave, Columbus, OH 43210, USA }
\\[3mm]
\textit{\small {}$^b$Department of Physics,
Stanford University\\ 382 Via Pueblo Mall, Stanford, CA 94305, USA }
\\[3mm]
\textit{\small {}$^\dagger$Kavli Institute for Theoretical Physics, \\ Santa Barbara, CA 93106, USA }
\end{center}

\vskip1cm

\begin{abstract}
In this paper we discuss the issues of supersymmetry breaking and
moduli stabilization within the context of $E_8 \otimes E_8$
heterotic orbifold constructions and, in particular, we focus on the
class of ``mini-landscape" models.  In the supersymmetric limit,
these models admit an effective low energy field theory with a
spectrum of states and dimensionless gauge and Yukawa couplings very
much like that of the MSSM.   These theories contain a non-Abelian
hidden gauge sector which generates a non-perturbative
superpotential leading to supersymmetry breaking and moduli
stabilization.   We demonstrate this effect in a simple model which
contains many of the features of the more general construction.  In
addition, we argue that once supersymmetry is broken in a restricted
sector of the theory, then all moduli are stabilized by supergravity
effects. Finally, we obtain the low energy superparticle spectrum
resulting from this simple model.

\end{abstract}

\end{titlepage}
\newpage

%==============================================================
%==============================================================
\section{Introduction}
\label{sec:intro}
%==============================================================
%==============================================================

String theory, as a candidate theory of all fundamental interactions
including gravity, is obliged to contain patterns consistent with
observation.  This includes the standard model gauge group and
particle content, as well as an extremely small cosmological
constant.  If one also assumes that nature contains a low energy
supersymmetry, one would want to find patterns which are
\textit{qualitatively} close to the MSSM.  An even more ambitious
goal would be to find a theory consistent with the standard model
spectrum of masses and a prediction for sparticle masses which can
be tested at the LHC. Some progress has been made to this end
starting from different directions~\cite{Cleaver:1998saa,
Braun:2005ux, Cvetic:2005bn, Buchmuller:2005jr, Buchmuller:2006ik,
Lebedev:2006kn, Kim:2007mt, Chen:2007px, Lebedev:2007hv,
Lebedev:2008un, Beasley:2008dc, Donagi:2008ca, Beasley:2008kw,
Donagi:2008kj, Blumenhagen:2008zz, Bourjaily:2009vf, Hayashi:2009ge,
Huh:2009nh, Marsano:2009gv}, i.e. free fermionic, orbifold or smooth
Calabi-Yau constructions of the heterotic string, intersecting
D-brane constructions in type II string,  and M or F theory
constructions. Much of this progress has benefited from the
requirement of an intermediate grand unified gauge symmetry which
naturally delivers the standard model particle spectrum.

In this paper we focus on the ``mini-landscape" of heterotic
orbifold constructions~\cite{Buchmuller:2005jr, Buchmuller:2006ik,
Lebedev:2006kn, Lebedev:2007hv, Lebedev:2008un}, which give several
models which pass a significant number of phenomenological
hurdles.\footnote{For reviews, see
\cite{Nilles:2008gq,Raby:2009dm}.}   These models have been analyzed
in the supersymmetric limit.  They contain an MSSM spectrum with
three families of quarks and leptons, one or more pairs of Higgs
doublets and an exact R parity.  In the orbifold limit, they also
contain a small number of vector-like exotics and extra $U(1)$ gauge
interactions felt by standard model particles.   These theories also
contain a large number of standard model singlet fields, some of
which are moduli, i.e.\thinspace blow up modes of the orbifold fixed points.
The superpotential for these orbifold theories can be calculated
order by order in powers of products of superfields. This is a
laborious task which is simplified by assuming that any term allowed
by string selection rules appears with an order one coefficient in
the superpotential. With this caveat it was shown that all
vector-like exotics and additional $U(1)$ gauge bosons acquire mass at
scales of order the string scale at supersymmetric minima satisfying
$ F_I = D_a = 0 $ for all chiral fields labeled by the index $I$ and
all gauge groups labeled by the index $a$.  In addition, the value
of the gauge couplings at the string scale and the effective Yukawa
couplings are determined by the presumed values of the vacuum
expectation values [VEVs] for moduli including the dilaton, $S$, the
bulk volume and complex structure moduli, $T_i, i = 1,2,3$ and $U$ and the
SM singlet fields containing the blow-up moduli
\cite{Nibbelink:2007pn,Nibbelink:2009sp}. Finally the theories also
contain a hidden sector non-Abelian gauge group with QCD-like chiral
matter. The problem which has yet to be addressed is the mechanism of
moduli stabilization and supersymmetry breaking in the
``mini-landscape" models.\footnote{For a preliminary analysis, see
\cite{Lebedev:2006tr}. Also moduli stabilization and supersymmetry
breaking in Type II string models and F theory constructions have
been considered in
\cite{Giddings:2001yu,Kachru:2003aw,Heckman:2008es,Marsano:2008jq,Heckman:2008qt}.}

In this paper we focus on the problem of moduli stabilization and
SUSY breaking in the context of heterotic orbifold models.  In
Section \ref{sec:general_structure} we summarize the general
structure of the K\"{a}hler and superpotential in heterotic orbifold
models.  The models have a perturbative superpotential satisfying
modular invariance constraints, an anomalous $U(1)_A$ gauge symmetry
with a dynamically generated Fayet-Illiopoulos $D$-term and a hidden
QCD-like non-Abelian gauge sector generating a non-perturbative
superpotential. In Section \ref{sec:one_gaugino_condensate} we
consider a simple model with a dilaton, $S$, one volume modulus,
$T$, and three standard model singlets.  The model has only one
gaugino condensate, as is the case for the ``benchmark models" of
the ``mini-landscape" \cite{Lebedev:2007hv}.  We obtain a `hybrid
KKLT' kind of superpotential that behaves like a single-condensate
for the dilaton $S$, but as a racetrack for the $T$ and, by
extension, also for the $U$ moduli; and an additional matter $F$
term, driven by the cancelation of an anomalous $U(1)_A$ $D$-term,
is the seed for successful up-lifting.  Previous analyses in the
literature have also used an anomalous $U(1)_A$ $D$-term in
coordination with other perturbative or non-perturbative terms in
the superpotential to accomplish SUSY breaking and
up-lifting~\cite{Binetruy:1996uv,Dvali:1996rj,Dvali:1997sr,Lalak:1997ar,deCarlos:2005kh,Choi:2006bh,Dudas:2006vc,Lalak:2007qd,Dudas:2007nz,Gallego:2008sv,Dudas:2008qf}.
We save a brief comparison of our work with some of these former
analyses for Section \ref{sec:one_gaugino_condensate}.  In Section
\ref{sec:moduli_stabilization} we discuss the other moduli and their
stabilization.  We conclude that a single gaugino condensate is
sufficient to break supersymmetry, stabilize all the moduli and
generate a de Sitter vacuum. Finally in Section \ref{sec:spectrum}
we evaluate the SUSY particle spectrum relevant for the LHC.  The
main results from this analysis are listed in Tables
\ref{tab:weak_obs}, \ref{tab:param_space}.

%==============================================================
%==============================================================
\section{General structure}
\label{sec:general_structure}
%==============================================================
%==============================================================
In this section we consider the supergravity limit of heterotic
orbifold models. However, we focus on the  ``mini-landscape" models
for definiteness. We discuss the general structure of the K\"{a}hler
potential, $\fk$, the superpotential, $\fw$, and gauge kinetic
function, $f_a$ for generic heterotic orbifold models. The
``mini-landscape" models are defined in terms of a \zsix
orbifold of the six internal dimensions of the ten dimensional
heterotic string.  The orbifold is described by a three dimensional
``twist'' vector $v$, which acts on the compact directions. We
define the compact directions in terms of complex coordinates:
 \begin{eqnarray} \nonumber
     Z_1 &\equiv& X_4 + iX_{5}, \\
     Z_2 &\equiv& X_6 + iX_{7}, \\ \nonumber
     Z_3 &\equiv& X_8 + iX_{9}.
 \end{eqnarray}
The twist is defined by the action $Z_i \rightarrow e^{2\pi i v_i}
Z_i$ for $i = 1,2,3$, and for \zsix we have $v =
\frac{1}{6} ( 1, 2, -3 )$ or a (60$^\circ$, 120$^\circ$,
180$^\circ$) rotation about the first, second and third torus,
respectively.  This defines the first twisted sector.   The second
and fourth twisted sectors are defined by twist vectors $2 v$ and $4
v$, respectively.   Note, the third torus is unaffected by this
twist.   In addition, for the third twisted sector, generated by the
twist vector $3 v$, the second torus is unaffected.   Finally the
fifth twisted sector, given by $5 v$ contains the $CP$ conjugate
states from the first twisted sector. Twisted sectors with
un-rotated tori contain $N = 2$ supersymmetric spectra.   This has
consequences for the non-perturbative superpotential discussed in
Section \ref{sec:gaugekinetic}.  Finally, these models have three
bulk volume moduli, $T_i, \ i=1,2,3$ and one bulk complex structure
modulus, $U$, for the third torus.

\subsection{Anomalous $U(1)_A$ and Fayet-Illiopoulos $D$-term}
\label{sec:anomalous}

The orbifold limit of the heterotic string has one anomalous
$U(1)_A$ symmetry. The dilaton superfield $S$, in fact, transforms
non-trivially under this symmetry.   Let $V_A, \ V_a$ be the gauge
superfields with gauge covariant field strengths, $W^\alpha_A,
W^\alpha_a$, of gauge groups, $U(1)_A, \ {\cal G}_a$, respectively.
The Lagrangian in the global limit is given in terms of a K\"{a}hler
potential
\cite{Green:1984sg,Fayet:1974jb,Dine:1987xk,Atick:1987gy,Dine:1987gj}
\be \fk =  - \log(S + {\overline S} - \delta_{GS} V_A)
  + \sum_{a} ({\overline Q}_a e^{V_a + 2 q_a V_A} Q_a + {\overline
{\tilde Q}}_a e^{- V_a + 2 \tilde q_a V_A} \tilde Q_a )
\label{eq:Kahler} \ee and a gauge kinetic superpotential \bea \fw =
& \frac{1}{2} [ \frac{S}{4} ( \sum_a k_a Tr W^\alpha_a W_{\alpha a}
+ k_A Tr W^\alpha_A W_{\alpha A} ) + h.c. ] . & \eea Note $q_a, \
\tilde q_a$ are the $U(1)_A$ charges of the `quark', $Q_a$, and
`anti-quark', $\tilde Q_a$, supermultiplets transforming under
${\cal G}_a$.

Under a $U(1)_A$ super-gauge transformation with parameter
$\Lambda$, one has
\bea \label{eq:u1A} \nonumber
\delta_A V_A &=& -i (\Lambda - \bar \Lambda)/2,  \\
\delta_A S   &=& -i \frac{\delta_{GS}}{2} \Lambda ,
\eea
and
\be
\delta_A \Phi = i q_\Phi \Lambda \Phi
\ee
for any charged multiplet $\Phi$. The
combination \be S + {\overline S} - \delta_{GS} V_A \ee is $U(1)_A$
invariant. $\delta_{GS}$ is the Green-Schwarz coefficient given by
\begin{equation} \delta_{GS} = 4 \frac{Tr Q_A}{192 \pi^2} =
\frac{(q_a + \tilde q_a) N_{f_a}}{4 \pi^2} \label{eq:GS}
\end{equation} where the middle term is for the
$U(1)_A-$gravity anomaly and the last term is for the $U(1)_A \times
({\cal G}_a)^2$ mixed anomaly.

The existence of an anomalous $U(1)_A$ has several interesting
consequences.  Due to the form of the K\"{a}hler potential (Eqn.
(\ref{eq:Kahler})) we obtain a Fayet-Illiopoulos $D$-term given by \be
\xi_A = \frac{\delta_{GS}}{2(S + {\overline S})} = - \frac{1}{2}
\delta_{GS} \
\partial_S \fk   \ee  with the $D$-term contribution to the scalar potential given by
\be V_D = \frac{1}{S + {\overline S}} \left( \sum_a X^A_a \partial_a
\fk \ \phi^a + \xi_A \right)^2  \ee where $X^A_a$ are Killing
vectors for $U(1)_A$. In addition, clearly the perturbative part of
the superpotential must be $U(1)_A$ invariant. But moreover, it
constrains the non-perturbative superpotential as well. In
particular, if the dilaton appears in the exponent,  the product
$e^{q_\Phi S} \Phi^{\delta_{GS}/2}$ is, and must also be, $U(1)_A$
invariant.

\subsection{Target space modular invariance}
\label{sec:modular}

In this section, we wish to present the modular dependence of the
gauge kinetic function, the K\"{a}hler potential, and of the
superpotential in as general a form as possible.  Most studies in
the past have worked with a universal $T$ modulus, and neglected the
effects of the $U$ moduli altogether.  Such a treatment is
warranted, for example, in the $\mathbb{Z}_3$ orbifolds where there
are no $U$ moduli.  If we want to work in the limit of a stringy
orbifold GUT \cite{Kobayashi:2004ya} which requires one of the $T$
moduli to be much larger than the others, or in the
\zsix orbifolds, however, it is impossible to treat all
of the $T$ and $U$ moduli on the same footing.

Consider the $SL(2,\mathbb{Z})$ modular transformations of $T$ and
$U$ given by
\cite{Witten:1985xb,Ferrara:1986qn,Cvetic:1988yw,Shapere:1988zv,Ferrara:1989bc,Lauer:1989ax,Chun:1989se,Ferrara:1989qb,Lauer:1990tm,Erler:1991ju,Erler:1991an,Stieberger:1992bj}\footnote{For
an excellent review with many references, see \cite{Bailin:1999nk}.}
\begin{equation}
 T \rightarrow \frac{aT-ib}{icT+d},\,\,\,ad-bc = 1,\,\,\,a,b,c,d\in\mathbb{Z},
\end{equation}
and
\begin{equation}
\log\left(T + \bar{T}\right) \rightarrow \log\left(\frac{T +
\bar{T}}{(icT + d)(-ic\bar{T} + d)}\right).
\end{equation}
The K\"{a}hler potential for moduli to zeroth order is given by:
\begin{eqnarray} \label{moduli_kahler0} \nonumber
\fk   &=& -\sum_{i=1}^{h_{(1,1)}} \log\left(T^i + \bar{T}^i\right) -
\sum_{j=1}^{h_{(2,1)}} \log\left(U^j + \bar{U}^j\right)  \\
 & = &- \sum_{i=1}^3 \log\left(T^i + \bar{T}^i\right) - \log\left(U + \bar{U}\right)
\end{eqnarray} where the last term applies to the ``mini-landscape" models, since in this case $h_{(1,1)} = 3, \
h_{(2,1)} = 1$.  Under the modular group, the K\"{a}hler potential
transforms as
\begin{equation}
\fk\rightarrow\fk + \sum_{i=1}^{h_{(1,1)}}\log |ic_iT^i + d_i|^2 + \sum_{j=1}^{h_{(2,1)}}
\log |ic_jU^j + d_j|^2 .
\end{equation}

The scalar potential $V$ is necessarily modular invariant.  We have
\be V = e^{\cal G} \left( {\cal G}_I {\cal G}^{I \bar J} {\cal
G}_{\bar J} - 3 \right) \ee   where  ${\cal G} = \fk + \log|\fw|^2$.
Hence for the scalar potential to be invariant under the modular
transformations, the superpotential must also transform as follows:
\begin{eqnarray} \nonumber \label{superpotenial_modular_transformation}
 \fw &\rightarrow& \prod_{i=1}^{h_{(1,1)}}\prod_{j=1}^{h_{(2,1)}}(ic_iT^i+d_i)^{-1}
 (ic_jU^j+d_j)^{-1}\fw, \\
 \bar{\fw} &\rightarrow& \prod_{i=1}^{h_{(1,1)}}\prod_{j=1}^{h_{(2,1)}}(-ic_i\bar{T}^i+d_i)^{-1}
 (-ic_j\bar{U}^j+d_j)^{-1}\bar{\fw}.
\end{eqnarray}
This can be guaranteed by appropriate powers of the Dedekind $\eta$
function multiplying terms in the superpotential.\footnote{These
terms arise as a consequence of world-sheet instantons in a string
calculation.  In fact, world sheet instantons typically result in
more general modular functions
\cite{Lauer:1989ax,Chun:1989se,Ferrara:1989qb,Lauer:1990tm,Erler:1991ju,Erler:1991an,Stieberger:1992bj}.}
This is due to the fact that under a modular transformation, we have
\begin{equation}
 \eta(T)\rightarrow(icT + d)^{1/2}\eta(T),
\end{equation}
up to a phase,  where \be \eta(T) = \exp(-\pi
T/12)\prod_{n=1}^\infty \left(1 - e^{- 2 \pi n T}\right).  \label{eq:etaT}
\ee

The transformation of both the matter fields and the superpotential
under the modular group fixes the modular dependence of the
interactions.  A field in the superpotential transforms as
\begin{equation}
 \Phi_I \rightarrow \Phi_I \prod_{i=1}^{h_{(1,1)}}\prod_{j=1}^{h_{(2,1)}}
 \left(ic_iT^i+d_i\right)^{-n_I^i}\left(ic_jU^j+d_j\right)^{-\ell_I^j}.
\end{equation}
The modular weights $n_I^i$ and $\ell_I^j$
\cite{Dixon:1989fj,Ibanez:1992hc} depend on the localization of the
matter fields on the orbifold.   For states $I$ in the $i$th
untwisted sector, i.e. those states with internal momentum in the
$i$th torus, we have $n_I^i = \ell_I^i = 1$, otherwise the weights
are 0.   For twisted sector states, we first define $\vec{\eta}(k)$,
which is related to the twisted sector $k (=1,\ldots,N-1)$ and the
orbifold twist vector $v$ by
\begin{equation}
\eta_i(k) \equiv kv_i \mod 1.
\end{equation}
Further, we require
\begin{equation}
 \sum_i \eta_i(k) \equiv 1.
\end{equation}
Then the modular weight of a state in the $k$th twisted sector is
given by
\begin{eqnarray} \label{modular_weight_1}
 n_I^i \equiv & (1-\eta^i(k)) + N^i - \bar{N}^i  & {\rm for} \;\; \eta_i(k) \neq 0 \\
 n_I^i \equiv &  N^i - \bar{N}^i & {\rm for} \;\;  \eta_i(k) = 0.  \nonumber
\end{eqnarray}
 The $N^i \ (\bar{N}^i)$ are
\textit{integer} oscillator numbers for left-moving oscillators
$\tilde{\alpha}^i$ ($\bar{\tilde{\alpha}}^{\bar{i}}$), respectively.
Similarly,
\begin{eqnarray}
 \ell_I^i \equiv & (1-\eta^i(k)) - N^i + \bar{N}^i & {\rm for} \;\;  \eta_i(k) \neq 0 \\
 \ell_I^i \equiv & - N^i + \bar{N}^i & {\rm for} \;\; \eta_i(k)  = 0.  \nonumber
\end{eqnarray}

In general, one can compute the superpotential to arbitrary order in
powers of superfields by a straightforward application of the string
selection rules \cite{Hamidi:1986vh, Casas:1991ac, Erler:1992gt,
patrick_phd}. One assumes that any term not forbidden by the string
selection rules appears with order one coefficient.  In practice,
even this becomes intractable quickly, and we must cut off the
procedure at some low, finite order.  More detailed calculations of
individual terms give coefficients dependent on volume moduli due to
string world sheet instantons.  In general the moduli dependence can
be obtained using the constraint of target space modular invariance.
Consider a superpotential term for the ``mini-landscape" models,
with three $T$ moduli and one $U$ modulus, of the form:
\begin{equation}
 \fw_3 = w_{IJK} \Phi_I \Phi_J \Phi_K.
\end{equation}
We assume that the fields $\Phi_{I,J,K}$ transform with modular
weights $n^i_{I,J,K}$ and $\ell^3_{I,J,K}$ under $T_i, \ i=1,2,3$
and $U$, respectively. Using the (net) transformation property of
the superpotential, and the transformation property of $\eta(T)$
under the modular group, we have (for non-universal moduli):
\be
\nonumber \label{t_depend}
 w_{IJK} \sim h_{IJK} \prod_{i=1}^3 \eta(T_i)^{\gamma_{T_i}} \eta(U)^{\gamma_{U}}
\ee where $ \gamma_{T_i} = -2(1-n_I^i-n_J^i-n_K^i) , \; \gamma_U =
-2(1-\ell_I^3-\ell_J^3-\ell_K^3)$.\footnote{Note,  the constants
$\gamma_{T_i}, \ \gamma_U$ can quite generally have either sign,
depending upon the modular weights of the fields at the particular
vertex.} This is easily generalized for higher order interaction
terms in the superpotential.   We see that the modular dependence of
the superpotential is rarely symmetric under interchange of the
$T_i$ or $U_i$.   Note, when minimizing the scalar potential we
shall use the approximation $\eta(T)^{\gamma_T} \approx e^{-bT}$
with $b = \pi \gamma_T/12$. (Recall, at large $T$, we have
$\log(\eta(T)) \approx - \pi T/12$.) This approximation misses the
physics near the self-dual point in the  potential, nevertheless, it
is typically a good approximation.

As a final note,  Wilson lines break the $SL(2, \mathbb{Z})$ modular
group down to a subgroup \cite{Love:1996sk} (see Appendix
\ref{app:diff_racetrack}). This has the effect of an additional
differentiation of the moduli as they appear in the superpotential.
In particular,  factors of $\eta(T_i)$ are replaced by factors of
$\eta(N T_i)$ or $\eta(T_i/N)$ for Wilson lines in $\mathbb{Z}_N$.
In summary, the different modular dependence of twisted sector
fields and the presence of Wilson lines leads quite generally to
anisotropic orbifolds \cite{Bailin:1994hu}.

\subsection{Gauge kinetic function and sigma model anomaly}
\label{sec:gaugekinetic}

To one loop, the string-derived gauge kinetic function is given by
\cite{Dixon:1990pc,Derendinger:1991hq,Ibanez:1991zv,Lust:1991yi,Ibanez:1992hc,Kaplunovsky:1995jw}
\begin{eqnarray} \nonumber \label{gauge_kinetic_one_loop}
f_a(S,T) &=& k_a S + \frac{1}{8\pi^2} \sum_{i=1}^{h_{(1,1)}}
\left(\alpha_a^i - k_a\delta_{\sigma}^i\right) \log \left(\eta(T^i)\right)^2 \\
&& + \frac{1}{8\pi^2} \sum_{j=1}^{h_{(2,1)}}
\left(\alpha_a^j - k_a\delta_{\sigma}^j\right) \log \left(\eta(U^j)\right)^2
\end{eqnarray}
where $k_a$ is the Ka\v{c}-Moody level of the group, which we will
normally take to be 1.  The constants $\alpha_a^i$ are model
dependent, and are defined as
\begin{equation} \nonumber
   \alpha_a^i \equiv \ell(\text{adj})- \sum_{\text{rep}_I} \ell_a(\text{rep}_I)(1+2n^i_I).
\end{equation}
$\ell(\text{adj})$ and $\ell_a(\text{rep}_I)$ are the Dynkin indices
of the adjoint representation and of the matter representation $I$
of the group $\fg_a$, respectively \cite{Slansky:1981yr} and $n^i_I$
are modular weights.\footnote{If $T_a^r$ are the generators of the
group $G_a$ in the representation $r$, then we have $Tr(T_a^r T_b^r)
= \ell_a(\text{rep}_r)\delta_{a b}$.}  The $\delta_{\sigma}^i$ terms
are necessary to cancel an anomaly in the underlying $\sigma$-model,
which induces a transformation in the dilaton field under the
modular group:
\begin{equation}
   S \rightarrow S + \frac{1}{8\pi^2} \sum_{i=1}^{h_{(1,1)}} \delta_{\sigma}^i
   \log\left(ic_iT_i + d_i\right) + \frac{1}{8\pi^2} \sum_{j=1}^{h_{(2,1)}}
   \delta_{\sigma}^i \log\left(ic_jU^j+ d_j\right).
\end{equation}

It is important to note that the factor  \be \left(\alpha_a^i -
k_a\delta_{\sigma}^i\right) \equiv \frac{b_a^{(N=2)}(i)}{|D|/|D_i|}
\ee  where $b_a^{(N=2)}(i)$ is the beta function coefficient for the
$i$th torus.  It is non-zero if and only if the $k$-th twisted
sector has an effective $N=2$ supersymmetry. Moreover this occurs only
when, in the $k$-th twisted sector, the $i$th torus is not rotated.
The factors $|D|, \ |D_i|$ are the degree of the twist group $D$ and
the little group $D_i$, which does not rotate the $i$th torus. For
example, for the ``mini-landscape" models with $D = \zsix$
we have $|D| = 6$ and $|D_2|= 2, \ |D_3| = 3$ since the little group
keeping the second (third) torus fixed is $\mathbb{Z}_2 \
(\mathbb{Z}_3)$.  The first torus is rotated in all twisted sectors.
Hence, the gauge kinetic function for the ``mini-landscape" models
is only a function of $T_2$ and $T_3$.

Taking into account the sigma model anomalies, the heterotic string
K\"{a}hler potential has the following form, where we have included
the loop corrections to the dilaton
\cite{Dixon:1990pc,Derendinger:1991hq}
\begin{eqnarray} \label{moduli_kahler} \nonumber
\fk &=& -\log\left(S + \bar{S} + \frac{1}{8\pi^2}\sum_{i=1}^{h_{(1,1)}}
 \delta_{\sigma}^i \log\left(T^i + \bar{T}^i\right) + \frac{1}{8\pi^2}
 \sum_{j=1}^{h_{(2,1)}} \delta_{\sigma}^j \log\left(U^j + \bar{U}^j\right)\right) \\
 &&-\sum_{i=1}^{h_{(1,1)}} \log\left(T^i + \bar{T}^i\right) -
 \sum_{j=1}^{h_{(2,1)}} \log\left(U^j + \bar{U}^j\right).
\end{eqnarray}
The first line of Eqn. (\ref{moduli_kahler}) is modular invariant by
itself, and one can redefine the dilaton, $Y$, such that
\begin{equation}
Y \equiv S + \bar{S} + \frac{1}{8\pi^2}\sum_{i=1}^{h_{(1,1)}} \delta_{\sigma}^i
\log\left(T^i + \bar{T}^i\right) + \frac{1}{8\pi^2}\sum_{j=1}^{h_{(1,2)}}
\delta_{\sigma}^j \log\left(U^j + \bar{U}^j\right),
\end{equation}
where $Y$ is invariant under the modular transformations.

\subsection{Non-perturbative superpotential}
\label{sec:NP}

In all ``mini-landscape" models \cite{Lebedev:2006tr}, and most
orbifold heterotic string constructions, there exists a hidden
sector with non-Abelian gauge interactions and vector-like matter
carrying hidden sector charge.   In the ``benchmark"
models \cite{Lebedev:2007hv} the hidden sector gauge group is $SU(4)$
with chiral matter in the $4 + \bar 4$ representation.

In this section let us consider a generic hidden sector with gauge
group $SU(N_{1})\otimes SU(N_2)\otimes U(1)_A$, where `$A$' stands
for anomalous.  There are $N_{f_1}$ and $N_{f_2}$ flavors of quarks
$Q_{1}$ and $Q_{2}$ in the fundamental representation (along with
anti-quarks $\tilde{Q}_1$ and $\tilde{Q}_2$, in the anti-fundamental
representations), as well as two singlet fields, called $\phi$ and
$\chi$.  The charge assignments are listed in Table
\ref{tab:GS_charge_assignments}.  We assume the existence of two
moduli, $S$ and $T$, which enter the non-perturbative superpotential
through the gauge kinetic function, namely $f=f(S,T)$.  The model
also allows for $T$ dependence in the Yukawa sector.

Non-perturbative effects generate a potential for the $S$ and $T$
moduli.  Gaugino condensation will generate a scale $\lsqcd$, which
is determined purely by the symmetries of the low energy theory:
\begin{equation} \label{gaugino_condensate_general}
   \Lambda_a(S,T) = e^{-\frac{8\pi^2}{\beta_a}f_a(S,T)},
\end{equation}
where $\beta_a = 3 N_a - N_{f_a}$ is the one loop beta function
coefficient of the theory. At tree level $f_a(S,T) = S$, however, we
include the possibility of threshold corrections which introduce a
dependence on the $T$ modulus
\cite{Dixon:1990pc,Derendinger:1991hq}.  We also find that $U(1)_A$
and modular invariance together dictate a very specific form for the
non-perturbative superpotential.

%=========================================================
\begin{table}
 \caption{\label{tab:GS_charge_assignments} Charge assignments for the
fields in a generic hidden sector. Flavor indices are suppressed.}
\centering
 \begin{tabular}{c|cccccc}
        &$\phi$     &$\chi$     &$Q_1$  &$Q_2$  &$\tilde{Q}_1$  &$\tilde{Q}_2$  \\
 \hline
 $\U1_A$    &-1     &$q_{\chi}$ &$q_1$  &$q_2$  &$\tilde{q}_1$  &$\tilde{q}_2$  \\
 $\SU{N_1}$ &1      &1      &$\Box$ &1  &$\bar{\Box}$   &1      \\
 $\SU{N_2}$ &1      &1      &1  &$\Box$ &1      &$\bar{\Box}$
 \end{tabular}
\end{table}
%=========================================================

In the ``mini-landscape" analysis the effective mass terms for the
vector-like exotics were evaluated.  They were given as a polynomial
in products of chiral MSSM singlet fields [chiral moduli].  It was
shown that all vector-like exotics obtain mass~\footnote{In fact,
one of the $SU(4)$ quark- anti-quark pairs remained massless in the
two ``benchmark" models.} when the chiral moduli obtain VEVs at
supersymmetric points in moduli space.   In our example let
us, for simplicity, take couplings between the quarks and the field
$\phi$ to be diagonal in flavor space. Mass terms of the form \be
\mathbb{M}_1(\phi,T) Q_{1}\tilde{Q}_{1} + \mathbb{M}_2(\phi,T)
Q_{2}\tilde{Q}_{2} \ee are dynamically generated when $\phi$
receives a non-zero VEV, which we will discuss below.  A key
assumption is that those mass terms are larger than the scale of
gaugino condensation, so that the quarks and anti-quarks may be
consistently integrated out.  If this can be accomplished, then one
can work in the pure gauge limit
\cite{deCarlos:1991gq}.\footnote{There is a check on the consistency
of this approach: at the end of the day, after calculating the VEVs
of the scalars, we can verify that the mass terms for the
quarks are indeed of the correct magnitude.}

Before we integrate out the meson fields, the non-perturbative
superpotential (plus quark masses) for $N_{f_a} < N_a$ is of the
form \cite{Affleck:1983mk}
\begin{equation} \label{full_np_w}
   \mathcal{W}_{\textsc{np}} = \sum_{a= 1,2} \left[ \mathbb{M}_a(\phi,T) Q_{a}\tilde{Q}_{a}
    + (N_a-N_{f_a})\left(\frac{\Lambda_a^{3N_a-N_{f_a}}}{\det Q_a
   \tilde{Q}_a}\right)^{\frac{1}{N_a-N_{f_a}}} \right],
\end{equation}
with $\mathbb{M}_a(\phi,T) = c_a e^{-b_a T}\phi^{q_a+\tilde{q}_a}$
where $c_a$ is a constant.  Note, given the charges for the fields
in Table \ref{tab:GS_charge_assignments} and using Eqns.
(\ref{eq:u1A}), (\ref{eq:GS}) and (\ref{gaugino_condensate_general}), one
sees that $\mathcal{W}_{\textsc{np}}$ is $U(1)_A$ invariant. The
K\"{a}hler potential for the hidden sector is assumed to be of the
form
\begin{eqnarray}
\fk & =  -\log(S + {\overline S}) - 3 \log(T + {\overline T}) + \alpha_\phi
{\overline \phi} e^{-2 V_A} \phi + \alpha_\chi {\overline \chi} e^{2 q_\chi V_A} \chi &  \\
& + \sum_{a = 1,2} \alpha_a ({\overline Q}_a e^{V_a + 2 q_a V_A} Q_a +
{\overline {\tilde Q}}_a e^{- V_a + 2 \tilde q_a V_A} \tilde Q_a  \nonumber
\end{eqnarray}
The quantities $\alpha_\phi, \alpha_\chi, \alpha_i$ are generally
functions of the modulus $T$, where the precise functional
dependence is fixed by the modular weights of the fields (see
Section \ref{sec:modular}).  $V_i$ and $V_A$ denote the vector
superfields associated with the gauge groups $\fg_i = SU(N_i)$ and
$U(1)_A$.

The determinant of the quark mass matrix is given by
\begin{equation} \label{quark_mass_matrix}
   \det\mathbb{M}_a(\phi,T) = \left(c_a e^{-b_a T}\phi^{q_a+\tilde{q}_a}\right)^{N_{f_a}}.
\end{equation}
We have taken the couplings between $\phi$ and the quarks to have
exponential dependence on the $T$ modulus, an ansatz which is
justified by modular invariance (see Section \ref{sec:modular}).
Inserting the meson equations of motion and Eqn.
(\ref{quark_mass_matrix}) into Eqn. (\ref{full_np_w}), we have
\begin{equation} \nonumber \label{racetrack}
   \mathcal{W}_{\textsc{np}} = \sum_{a = 1,2} \left[ N_a \left(c_a e^{-b_a T}
   \phi^{q_a+\tilde{q}_a}\right)^{\frac{N_{f_a}}{N_a}}
   \left[\Lambda_a(S,T)\right]^{\frac{3N_a-N_{f_a}}{N_a}} \right].
\end{equation}

Note that the transformation of the superpotential under the modular
group in Eqn. (\ref{superpotenial_modular_transformation}) also
requires that the (non-perturbative) superpotential obey
\begin{equation}
   \fw_{\textsc{np}} \rightarrow \prod_{i=1}^{h_{(1,1)}}\prod_{j=1}^{h_{(2,1)}}
   (ic_iT^i+d_i)^{-1}(ic_jU^j+d_j)^{-1}\fw_{\textsc{np}}.
\end{equation}
Because the non-perturbative lagrangian must be invariant under all
of the symmetries of the underlying string theory, it must be that
\cite{Font:1990nt,Ferrara:1990ei,Nilles:1990jv,Casas:1990qi,Lust:1991yi,deCarlos:1992da}:
\begin{equation} \label{duality_invariant_gaugino_condensate}
\fw_{\textsc{np}} \equiv A \times e^{-a S}\prod_{i=1}^{h_{(1,1)}}
   \prod_{j=1}^{h_{(2,1)}} \left(\eta(T^i)\right)^{-2+\frac{3}{4\pi^2 \beta}\delta^i_{\sigma}}
   \left(\eta(U^j)\right)^{-2+\frac{3}{4\pi^2 \beta}\delta^j_{\sigma}}
\end{equation}  where $a \equiv \frac{24 \pi^2}{\beta}$ and
$\beta = 3 \ell(\text{adj}) - \sum_{I} \ell(\text{rep}_I)$ is the
one-loop beta function coefficient, and $A$ is generally a function of
the chiral matter fields appearing in $\mathbb{M}$. This, coupled with the one loop
gauge kinetic function in Eqn. (\ref{gauge_kinetic_one_loop}), gives
the heterotic generalization of the Racetrack superpotential.

In the following Section \ref{sec:one_gaugino_condensate}, we
construct a simple model using the qualitative features outlined in
this section. This model is novel because it requires only one
non-Abelian gauge group to stabilize moduli and give a de Sitter
vacuum. We have also constructed two condensate models,
however, the literature already contains several examples of the
``racetrack" in regards to stabilization of $S$ and $T$ moduli.
Moreover in the ``mini-landscape" models, whose features we are
seeking to reproduce, there are many examples of hidden sectors
containing a single non-Abelian gauge group \cite{Lebedev:2006tr},
while there are no examples with multiple hidden sectors.

%==============================================================
%==============================================================
\section{Moduli stabilization and supersymmetry\\ breaking in the bulk}
\label{sec:one_gaugino_condensate}
%==============================================================
%==============================================================
In this section we construct a simple, generic heterotic orbifold
model which captures many of the features discussed in Section
\ref{sec:general_structure}.   In particular, it is a single gaugino
condensate model with the following fields - dilaton ($S$), modulus
($T$) and MSSM singlets ($\phi_1, \phi_2, \chi$).   The model has
one anomalous $\U1_A$ with the singlet charges given by ($q_{\phi_1}
= -2, q_{\phi_2} = -9, q_\chi = 20$). The K\"{a}hler and
superpotential are given by~\footnote{The coefficient $A$ (Eqn.
(\ref{eqn:A})) is an implicit function of all other non-vanishing
chiral singlet VEVs which would be necessary to satisfy the modular
invariance constraints, i.e. $A = A(\langle \phi_I \rangle)$.  If
one re-scales the $U(1)_A$ charges, $q_{\phi_i}, q_\chi \rightarrow
q_{\phi_i}/r, q_\chi/r$, then the $U(1)_A$ constraint is satisfied
with $r = 15 p$ (assuming no additional singlets in $A$).  Otherwise
we may let $r$ and $p$ be independent. This re-scaling does not
affect our analysis, since the vacuum value of the $\phi_i, \chi$
term in the superpotential vanishes. \label{footnote} }
\begin{eqnarray} \label{model_Kahler_potential}
\fk = & -\log[S + \bar S] - 3 \log[T + {\overline T}] + {\overline \phi}_1 \phi_1 +
{\overline \phi}_2 \phi_2 + {\overline \chi} \chi & \\ \label{model_superpotential}
\fw = & e^{-b T} (w_0 + \chi  (\phi_1^{10} + \lambda \phi_1 \phi_2^2))
+ A \ \phi_2^p \ e^{-a S - b_2 T}. &  \label{eqn:A}
\end{eqnarray}
In addition, there is an anomalous $U(1)_A$ $D$-term given by
\begin{eqnarray}
D_A = & 20 {\overline \chi} \chi - 2 {\overline \phi}_1 \phi_1 - 9 {\overline \phi}_2 \phi_2 -
\frac{1}{2} \delta_{GS} \ \partial_S \fk  &
\end{eqnarray}
with $\delta_{GS} =  \frac{(q + \tilde q) N_{f}}{4 \pi^2} = N_f/(4
\pi^2)$.

In the absence of the non-perturbative term (with coefficient $A$)
the theory has a supersymmetric minimum with $\langle \chi \rangle =
\langle \phi_1 \rangle = 0$ and $\langle \phi_2 \rangle \neq 0$ and
arbitrary.  This property mirrors the situation in the
``mini-landscape" models where supersymmetric vacua have been found
in the limit that all non-perturbative effects are neglected.  We
have also added a constant $w_0 = w_0(\langle \phi_I \rangle)$ which
is expected to be generated (in the ``mini-landscape" models) at
high order in the product of chiral moduli due to the explicit
breaking of an accidental $R$ symmetry which exists at lower
orders~\cite{Kappl:2008ie}.\footnote{The fields entering $w_0$ have
string scale mass.}  The $T$ dependence in the superpotential is
designed to take into account, in a qualitative way, the modular
invariance constraints of Section \ref{sec:modular}. We have
included only one $T$ modulus, assuming that the others can be
stabilized near the self-dual point
\cite{Font:1990nt,Cvetic:1991qm}. Moreover, as argued earlier, the
$T_i$ and $U$ moduli enter the superpotential in different ways (see
Section \ref{sec:modular}). This leads to modular invariant
solutions which are typically
anisotropic~\cite{Bailin:1994hu}.\footnote{Note, we have chosen to
keep the form of the K\"{a}hler potential for this single $T$
modulus with the factor of 3, so as to maintain the approximate
no-scale behavior.}

Note, that the structure, $\fw \sim w_0 e^{- b T} + \phi_2 \ e^{- a
S - b_2 T}$ gives us the crucial progress\footnote{Note, the
constants $b, \ b_2$ can have either sign. For the case with $b, \
b_2 > 0$ the superpotential for $T$ is racetrack-like.  However for
$b, \ b_2 < 0$ the scalar potential for $T$ diverges as $T$ goes to
zero or infinity and compactification is guaranteed
\cite{Font:1990nt,Cvetic:1991qm}.} -
\begin{enumerate}[i.)]
\item  a `hybrid KKLT' kind of superpotential that behaves like a
single-condensate for the dilaton $S$, but as a racetrack for the
$T$ and, by extension, also for the $U$ moduli; and
\item an additional matter $F_{\phi_2}$ term driven by the
cancelation of the anomalous $U(1)_A$ $D$-term seeds SUSY breaking
with successful uplifing.
\end{enumerate}

The constant $b$ is fixed by modular invariance constraints. In
general the two terms in the perturbative superpotential would have
different $T$ dependence.   We have found solutions for this case as
well. This is possible since the VEV of the $\chi$ term in the
superpotential vanishes.  The second term (proportional to $A$)
represents the non-perturbative contribution of one gaugino
condensate.  The constants $a = 24 \pi^2/\beta, \ b_2$ and $p$
depend on the size of the gauge group, the number of flavors and the
coefficient of the one-loop beta function for the effective $N=2$
supersymmetry of the torus $T$. For the ``mini-landscape" models,
this would be either $T_2$ or $T_3$. Finally, the coefficient of the
exponential factor of the dilaton $S$ is taken to be $A \ \phi_2^p$.
This represents the effective hidden sector quark mass term, which
in this case is proportional to a power of the chiral singlet
$\phi_2$. In a more general case, it would be a polynomial in powers
of chiral moduli.\footnote{Holomorphic gauge invariant monomials
span the moduli space of supersymmetric vacua. One such monomial is
necessary to cancel the Fayet-Illiopoulos $D$-term (see Appendix
\ref{app:HIM}).} The exponent $p$ depends in general on the size of
the gauge group, the number of flavors and the power that the field
$\phi_2$ appears in the effective quark mass term.

We have performed a numerical evaluation of the scalar potential
with the following input parameters.  We take hidden sector gauge
group $SU(N)$ with $N = 5, \ N_f = 3$ and $a = 8
\pi^2/N$.\footnote{We have also found solutions for the case with $N
= 4, \ N_f = 7$ which is closer to the ``mini-landscape" benchmark
models.  Note, when $N_f > N$ we may still use the same formalism,
since we assume that all the $Q, \tilde Q$s get mass much above the
effective QCD scale.} For the other input values we have considered
five different possibilities given in Table
\ref{tab:onecond.input}.\footnote{Note the parameter relation $r =
15 p$ in Table \ref{tab:onecond.input} is derived using $U(1)_A$
invariance and the assumption that no other fields with
non-vanishing $U(1)_A$ charge enter into the effective mass matrix
for hidden sector quarks.  We have also allowed for two cases where
this relation is not satisfied.} We find that supersymmetry
breaking, moduli stabilization and up-lifting is a direct
consequence of adding the non-perturbative superpotential term.

In our analysis we use the scalar potential $V$ given by  \be V =
e^K (\sum_{i = 1}^5\ \sum_{j = 1}^5 \left[ F_{\Phi_i}\
\overline{F_{\Phi_j}}\ \fk^{-1}_{i, j} - 3 |W|^2 \right]) +
\frac{D_A^2}{(S + \bar S)} + \Delta V_{CW}[\Phi_i,
\overline{\Phi}_i] \ee where $\Phi_{i,j} = \{ S, T, \chi, \phi_1,
\phi_2 \}$ and $F_{\Phi_i} \equiv \partial_{\Phi_i} \fw +
(\partial_{\Phi_i} \fk) \fw$.  The first two terms are the tree
level supergravity potential.  The last term is a one loop
correction which affects the vacuum energy and $D$ term
contribution.

The one loop Coleman-Weinberg potential is in general given by \be
\Delta V_{CW} = \frac{1}{32 \pi^2} Str(M^2) \Lambda^2 + \frac{1}{64
\pi^2} Str(M^4 \log[\frac{M^2}{\Lambda^2}])  \ee  with the mass
matrix $M$ given by $M = M(\Phi_i)$ and $\Lambda$ is the relevant
cut-off in the problem.  We take $\Lambda = \mstring \sim 10^{17}$
GeV.

We have not evaluated the full one loop correction.  Instead we use
the approximate formula \be \Delta V_{CW}[\phi_2, \overline{\phi_2}]
= \frac{\lambda^2 \ F_2^2 \ |\phi_2|^2}{8 \pi^2} \left(\log[R
(\lambda |\phi_2|^2)^2] + 3/2\right) + \order{\Lambda^2} \ee where
$F_2 = \langle F_{\phi_2} \rangle$ is obtained self-consistently and
all dimensionful quantities are expressed in Planck units.  This one
loop expression results from the $\chi, \ \phi_1$ contributions to
the Coleman-Weinberg formula.   The term quadratic in the cut-off is
naturally proportional to the number of chiral multiplets in the
theory and could be expected to contribute a small amount to the
vacuum energy, of order a few percent times $m_{3/2}^2 M_{pl}^2$. We
will discuss this contribution later, after finding the minima of
the potential.  Finally, note that the parameters $\lambda, \ R$ in
Table \ref{tab:onecond.input} might both be expected to be
significantly greater than one when written in Planck units.   This
is because the scale of the effective higher dimensional operator
with coefficient $\lambda$ in Eqn. \ref{eqn:A} is most likely set by
some value between $M_{Pl}$ and $M_{string}$ and the cut-off scale
for the one loop calculation (which determines the constant $R$) is
the string scale and not $M_{Pl}$.

\begin{table}[ht]
\renewcommand\arraystretch{1.2}
\begin{center}
\begin{tabular}{|c|c|c|c|c|c|c|c|c|}
\hline
  Case &    $b$ & $b_2$     & $\lambda$ & $R$ & $p $ & $r$ & $A$ & $w_0$  \\
\hline
1 & $\pi/50$   & $3\pi/2$  & $33$  & $10$  & $2/5$ & $15p$ & $160$ & $8 \times 10^{-15}$ \\
\hline
2 & $8/125$    & $3\pi/2$  & $0$   &  $5$  & $2/5$ & $15p$ & $30$  & $42 \times 10^{-16}$ \\
\hline
3 & $1/16$ & $29\pi/20$    & $38$  & $10$  & $2/5$ & $15p$ & $90$  & $6 \times 10^{-15}$ \\
\hline
4 & $-\pi/120$& $-\pi/40$     & $40$  & $64$  & $2/3$ & $1$   & $1/10$& $-5 \times 10^{-15}$ \\
\hline
5 & $-\pi/250$ & $-\pi/100$    & $25$  & $16$  & $1$   & $10/3$& $7/5$ & $-7 \times 10^{-15}$ \\
\hline
\end{tabular}
\caption{\label{tab:onecond.input}Input values for the
superpotential parameters for three different cases.  Case 2 has a
vanishing one loop correction for $\phi_2$.}
\end{center}
\end{table}

\begin{table}[ht]
\renewcommand\arraystretch{1.2}
\centering
\noindent\makebox[\textwidth]{ \scalebox{0.75}{
\begin{tabular}{|c|c|c|c|c|c|}
\hline
 &  Case 1 & Case 2  & Case 3 & Case 4 & Case 5\\
\hline
$\vev{s}$ & $2.2$ & $2.2$ & $2.1$ & $2.1$ & $2.2$ \\
\hline
$\vev{t}$ & $1.2$ & $1.1$ & $1.6$ & $1.1$ & $1.1$ \\
\hline
$\vev{\sigma}$ & $1.0$ & $1.0$ & $1.0$ & $0.0$ & $0.0$ \\
\hline
$\vev{\phi_2}$ & $0.08$ & $0.08$ & $0.08$ & $0.03$ & $0.06$ \\
\hline
$F_S$ & $2.8 \times 10^{-16}$ & $1.3 \times 10^{-16}$ & $2.7 \times 10^{-16}$ & $1.1 \times 10^{-16}$ & $8.0\times10^{-17}$\\
\hline
$F_T$ & $-8.7 \times 10^{-15}$ & $-5.1 \times 10^{-15}$ & $-5.0 \times 10^{-15}$ & $6.7 \times 10^{-15}$ & $9.1\times10^{-15}$\\
\hline
$F_{\phi_2}$ & $-9.2\times 10^{-17}$ & $-4.5\times 10^{-17}$ & $-8.9\times 10^{-17}$ & $1.3\times10^{-15}$ & $1.3\times10^{-15}$ \\
\hline
$D_A $ & $4.4\times 10^{-31}$ & $1.0\times 10^{-32}$ &  $5.9\times 10^{-31}$ & $-3.8\times 10^{-31}$ & $-4.8\times10^{-32}$  \\
\hline
$D_A / m_{3/2}^2$ & $0.6$ & $0.03$ & $2.7$  & $-0.7$ & $-0.05$ \\
\hline
$V_0/(3 m_{3/2}^2)$ & $-0.02 $ & $-0.01 $  &  $-0.02$ & $-0.03$ & $-0.02$\\
\hline
$m_{3/2}$ & $2.2$ TeV & $1.4$ TeV  &  $1.1$ TeV & $1.8$ TeV & $2.4$ TeV \\
\hline
\end{tabular}
}}
\caption{\label{tab:onecond.result} The values for field VEVs and
soft SUSY breaking parameters at the minimum of the scalar potential.
Note $F_\Phi \equiv \partial_\Phi \fw + (\partial_\Phi \fk) \fw$.}
\end{table}

\begin{figure}[t!]
\centering
\subfigure[The scalar potential in Case 2 for ${\rm Re}\,T$, with $b_i > 0$.]
{\includegraphics[scale=0.8]{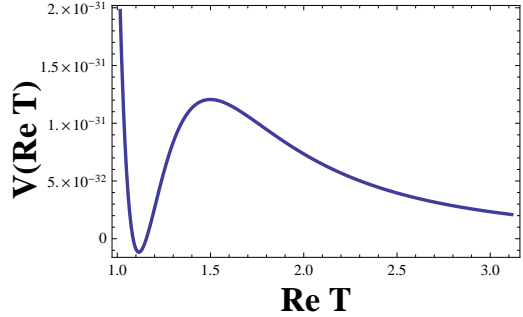} \label{fig:pos_b} }
\subfigure[The scalar potential in Case 4 for ${\rm Re}\,T$, with $b_i < 0$.]
{\includegraphics[scale=0.8]{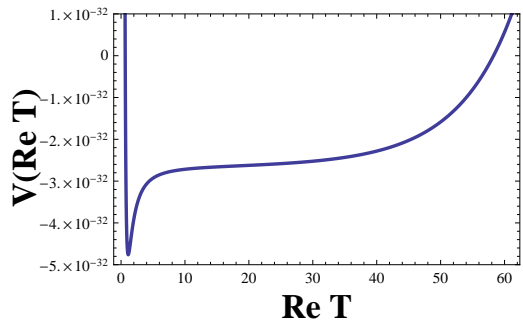} \label{fig:neg_b} }
\label{fig:potentials}
\caption[]{As ${\rm Re}\,T \rightarrow \infty$, the potential for
$b_i > 0$ mimics a Racetrack, which can be seen from Eqn.
(\ref{model_superpotential}), for example.  In the case where
$b_i < 0$, however, the potential exhibits a different asymptotic
behavior.  As ${\rm Re}\,T \rightarrow \infty$ the potential
diverges, which means that theory is forced to be compactified
\cite{Font:1990nt,Cvetic:1991qm}. }
\end{figure}

\begin{figure}[t!]
\centering
{\includegraphics[scale=0.8]{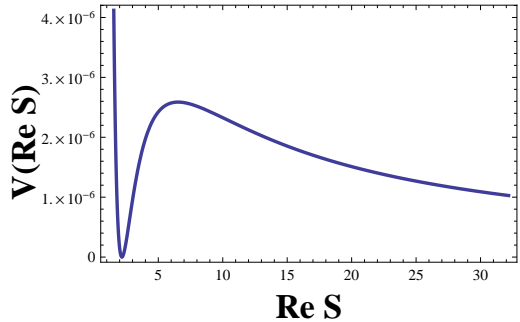}  }
\caption[]{The scalar potential in the \re{S} direction for Case 2. \label{fig:v_of_s}}
\end{figure}

In all cases we find a meta-stable minimum with all (except for two
massless modes) fields massive of \order{\text{TeV}} or larger.
Supersymmetry is broken at the minimum with values given in Table
\ref{tab:onecond.result}. Note $\re{S} \sim 2.2$ and $\re{T}$ ranges
between 1.1 and 1.6.  The moduli $\chi, \ \phi_1$ are stabilized at
their global minima $\phi_1 = \chi = 0 $ with $F_\chi = F_{\phi_1} =
0$ in all cases.  The modulus $\sigma = \im{S}$ is stabilized at
$\sigma \approx 1$ in the racetrack cases 1,2,and 3. This value
enforces a relative negative sign between the two terms dependent on
$\re{T}$. We plot the scalar potential $V$ in the \re{T} direction
for case 2 ($b, \ b_2 > 0$) (Fig. \ref{fig:pos_b}) and for case 4
($b, \ b_2 < 0$) (Fig. \ref{fig:neg_b}).  Note the potential as a
function of \re{S} is qualitatively the same for both cases (Fig.
\ref{fig:v_of_s}).

%=======================================================================
%=======================================================================

%=======================================================================
%=======================================================================

%=======================================================================
\begin{figure}[t!]
\centering
\includegraphics[scale=0.8]{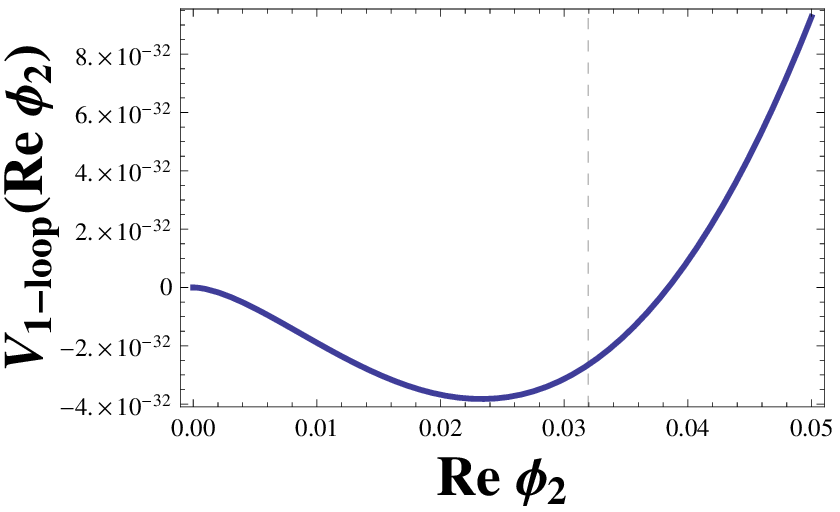}
\label{fig:one_loop_potential}
\caption{The one loop Coleman-Weinberg potential (Case 4) for $\phi_2$.
The dashed line represents the VEV of $\phi_2$ in the minimum of the
\textit{full} potential.}
\end{figure}
%=======================================================================

At the meta-stable minimum of the scalar potential we find a vacuum
energy which is slightly negative, i.e. of order $(-0.03 \; {\rm to}
\; -0.01)\times 3 m_{3/2}^2 M_{Pl}^2$ (see Table
\ref{tab:onecond.result}). Note, however, one loop radiative
corrections to the vacuum energy are of order $(N_T \ m_{3/2}^2
M_S^2/16 \pi^2)$ ,where $N_T$ is the total number of chiral
multiplets~\cite{Ferrara:1994kg} and we have assumed a cut-off at
the string scale $M_S$. With typical values $N_T \sim \order{300}$
and $M_S/M_{Pl} \sim 0.1$, this can easily lift the vacuum energy
the rest of the way to give a small positive effective cosmological
constant which is thus a meta-stable local dS minimum.  Note that
the constants $\lambda, \ R$ have also been used to adjust the value
of the cosmological constant as well as, and more importantly for
LHC phenomenology, the value of $D_A$ (see Fig.
\ref{fig:one_loop_potential}).

The two massless fields can be seen as the result of two $U(1)$
symmetries; the first is a $U(1)_R$ symmetry and the second is
associated with the anomalous $U(1)_A$.  The $U(1)_R$ is likely
generic (but approximate), since even the ``constant" superpotential
term needed to obtain a small cosmological constant necessarily
comes with $\eta(T)$ moduli dependence.   Since we have approximated
$\eta(T) \sim  \exp(- \pi T/12)$ by the first term in the series
expansion (Eqn. \ref{eq:etaT}), the symmetry is exact.  However
higher order terms in the expansion necessarily break the $U(1)_R$
symmetry. The $U(1)_A$ symmetry is gauged.

One can express the fields $S,T,$ and $\phi_2$ in the following
basis\footnote{The fields $\chi$ and $\phi_1$ cannot be expressed in
polar coordinates as they receive zero VEV, and cannot be
canonically normalized in this basis.}:
\begin{eqnarray} \nonumber
   S &\equiv& s + i\sigma, \\
   T &\equiv& t+i\tau, \\ \nonumber
   \phi_2 &\equiv& \varphi_2 \ e^{i\theta_2}.
\end{eqnarray}
The transformation properties of the fields $\sigma, \tau$ and
$\theta_2$ under the two \U1's are given by
\begin{eqnarray} \nonumber
   U(1)_R &:& \left\{ \begin{array}{rcl} \tau & \rightarrow & \tau + c \\ \sigma &\rightarrow& \sigma + \frac{-b_2 + b}{a} c \end{array}  \right. , \\
   U(1)_A &:& \left\{ \begin{array}{rcl} \theta & \rightarrow & \theta -\frac{9}{r}c' \\ \sigma &\rightarrow& \sigma - \frac{9p}{a\cdot r} c' \end{array}  \right. ,
\end{eqnarray}
where $c,c'$ are arbitrary constants and for the definition of $r$
see footnote \ref{footnote}. The corresponding Nambu-Goldstone (NG)
bosons are given by
\begin{eqnarray} \nonumber
   \chi^1_{\textsc{ng}} &=& \frac{a}{-b_2 + b} \sigma + \tau, \\
   \chi^2_{\textsc{ng}} &=& \tilde{N}\left(-\sigma + \frac{-b_2 + b}{a}\tau\right) + \frac{1}{p}\theta_2,
\end{eqnarray}
where $\tilde{N}$ is a normalization factor. One can then calculate
the mass matrix in the $\sigma-\tau-\theta_2$ basis and find two
zero eigenvalues (as expected) and one non-zero eigenvalue.  The two
NG modes, in all cases, can be shown to be linear combinations of
the two eigenvectors of the two massless states.  The $\U1_A$ NG
boson is eaten by the $\U1_A$ gauge boson, while the $\U1_R$
pseudo-NG boson remains as an ``invisible axion"
\cite{Nilles:1981py}.  The $U(1)_R$ symmetry is non-perturbatively
broken (by world-sheet instantons) at a scale of order \be  \langle
e^{\fk/2} \fw \ e^{- \pi T} \rangle \approx m_{3/2} \langle  e^{-
\pi T} \rangle \sim 0.02 \ m_{3/2} \ee in Planck units, resulting in
an ``axion" mass of order 10 GeV and decay constant of order
$\mpl$.\footnote{In addition, the heterotic orbifold models might
very well have the standard invisible axion \cite{Choi:2009jt}.}

Before discussing the rest of the moduli, in a more complete string
model,  and how they would be stabilized or the LHC phenomenology of
the mini-version of the mini-landscape models, it is worth comparing
our analysis with some previous discussions in the literature.

In a series of two papers by Dvali and Pomarol
\cite{Dvali:1996rj,Dvali:1997sr}, the authors consider an anomalous
\U1 with two charged singlet fields.  The $D$ term is given
by\footnote{We refer to the anomalous $U(1)$ as $U(1)_A$ and not
$U(1)_X$, as in the papers referenced below.}
\begin{equation}
   D_A = q |\phi_+|^2 - |q_-|^2 + \xi
\end{equation}
The gauge invariant superpotential is
\begin{equation}
   \fw = m \phi_+ \phi_-,
\end{equation}
where $m$ has some charge under $\U1_A$.  They suggest a few
different ways to generate $m$. The first is with some high power of
one of the $\phi$ fields:
\begin{equation}
   \fw \sim \phi_-^q \phi_+ \Rightarrow m \equiv \vev{\phi_-}^{q-1}
\end{equation}
The second is by giving the $\phi$ a coupling to some quarks from a
SUSY QCD theory that becomes strongly coupled.  The scale,
$\Lambda_{SQCD}$ then serves as the mass term in the superpotential.
They do not, however, consider dilaton dependence, and their $D$
term is static, not dynamic.  They also work in the global SUSY
limit, so they do not consider up-lifting.

In a paper by Binetruy and Dudas \cite{Binetruy:1996uv}, the authors
assume that $S$ can be stabilized at some finite value $S_0$,
possibly through some extra $S$ dependent term in the superpotential
and they assume that $F_S(S_0) = 0$.  In their setup, they have an
anomalous $\U1$, some charged singlets, and some hidden sector SQCD
with matter.  The singlets couple to matter, and SQCD becomes
strongly coupled, generating a scale, just as in our analysis. Since
they are working in the global SUSY limit, they are not concerned
with up-lifting.

Lalak \cite{Lalak:1997ar} considers several types of models with an
anomalous \U1, some charged singlets, and some coupling to the
dilaton $S$.  In the last section, he considers superpotentials with
an exponential dependence on $S$.  He then assumes that $S_0$ is a
(globally) supersymmetric minimum of the potential. Also, working in
global SUSY, he does not address up-lifting.

In a paper by Dudas and Mambrini \cite{Dudas:2006vc}, the authors
consider one modulus, one singlet field, and an $SU(N)$ with one
flavor of quarks.  The $SU(N)$ becomes strongly coupled, and the
superpotential and K\"{a}hler potential look like: \bea \fw = w_0 +
(c / X^{2}) e^{-a T} + m \phi^q X \\ \fk = -3 \log( T + \bar T -
|X|^2 - |\phi|^2 ) \eea where $X$ is the meson field and $\phi$ is
the singlet.  Note, the modulus appearing in the exponent is $T$,
not $S$.  They find that the only consistent minimum with
approximately zero cosmological constant requires $m_{3/2} \sim
\xi$.  So either the gravitino mass is of order the GUT scale or for
the gravitino mass of order a TeV, the meson charge must satisfy $q
\sim 10^{-8}$.

In a paper by Dudas et al. \cite{Dudas:2007nz}, the authors consider
a single modulus and two singlet fields:
\begin{eqnarray}
   D_A &=& |\phi_+|^2 - |\phi_-|^2 + \xi, \\
   \fw &=& w_0 + m\phi_+\phi_- + a \phi_-^q e^{-bT}.
\end{eqnarray}
They do not discuss the origin of the constant $w_0$.  They suggest
that $m$ might come from non-perturbative effects. Note the latter
is crucial, since $m$ affects the up-lifting of the scalar
potential. They are also interested in large volume
compactifications, as $t\equiv \re{T}\approx 60$. Given their SUSY
breaking scheme, they go on to look at the low energy spectrum.
However, they neglect the $D$ term contributions to the soft masses,
claiming that there are only two possibilities for the low energy
physics:
\begin{itemize}
   \item Because $\xi > 0$, some SM quarks and leptons carry
   positive $\U1_A$ charges.  This leads to scalar masses (for
   them) of around 100 TeV, and may give an unstable low energy
   spectrum.
   \item All SM quarks and leptons are neutral under $\U1_X$.
   This implies that there should be more matter that is charged
   under the MSSM and $\U1_A$.
\end{itemize}
It seems that they have missed an important possibility, namely that
matter in the MSSM appears with $\U1_A$ charges of both signs.  This
actually seems to be the generic case, at least in the
mini-landscape models.

The last paper we consider, by Gallego and Serone
\cite{Gallego:2008sv}, contains an analysis which is possibly most
similar to that in this paper.  There are however two major
differences.  If one neglects all non-perturbative dependence on the
dilaton and K\"{a}hler moduli, then their superpotential is of the
form $\fw \supset \phi^q \chi$ and the D term is given by $D_A = q
|\chi|^2 - |\phi|^2 + \xi$.  Hence the model does not have a
supersymmetric minimum in the global limit, due to a conflict
between $F_\chi = 0$ and $D_A = 0$.   However in oure model (Eqn.
\ref{eqn:A}) there is a supersymmetric solution when
non-perturbative effects are ignored. Finally,  the authors were not
able to find a supersymmetry breaking solution, like ours, with just
one hidden non-Abelian gauge sector.

As an aside, we note that Casas et al. \cite{Casas:1990qi} study a
similar problem of moduli stabilization and SUSY breaking, but
without the anomalous \U1. However,  their model is very different
from ours, but they do include the one loop Coleman-Weinberg
corrections.

%==============================================================
%==============================================================
\section{Moduli stabilization continued - the twisted sector and blow-up moduli}
\label{sec:moduli_stabilization}
%==============================================================
%==============================================================

In our discussion above we considered a simple model which is
representative of heterotic orbifold models.   Our simple model had
only a few moduli, i.e. the dilaton, $S$,  a volume modulus, $T$,
and three chiral singlet `moduli', $\chi, \ \phi_1, \ \phi_2$. Any
heterotic orbifold construction, on the other hand, will have
several volume and complex structure moduli and, of order 50 to 100
chiral singlet moduli.   The superpotential for the chiral singlet
moduli is obtained as a polynomial product of holomorphic gauge
invariant monomials which typically contain hundreds of terms at
each order (with the number of terms increasing with the order). In
the ``mini-landscape" analysis, supersymmetric vacua satisfying $F =
D = 0$ constraints to sixth order in chiral singlet moduli could be found.
Although there are many flat
directions in moduli space,  the anomalous $D$-term fixes at least one
holomorphic gauge invariant monomial to have a large value.   Our
simple model expressed this fact with the chiral singlets  $\chi, \
\phi_1, \ \phi_2$, where the VEVs were fixed by the global SUSY
minimum with $\langle \phi_2 \rangle$ fixed by the $U(1)_A$ $D$-term.

In addition to the non-Abelian hidden gauge sector considered in the
simple model, a generic orbifold vacuum also has additional $U(1)$
gauge interactions and vector-like exotics which obtain mass
proportional to chiral singlet VEVs. Some of these singlets are
assumed to get large VEVs (of order the string scale). These are the
ones giving mass to the extra $U(1)$ gauge sector and vector-like
exotics. These same VEVs generate non-trivial Yukawa couplings for
quarks and leptons.  Moreover, there are chiral singlets which get zero
VEVs, such as $\chi$ and $\phi_1$.  For example, in the
``mini-landscape" benchmark model 1, the electroweak Higgs $\mu$
term is zero in the supersymmetric limit. The question arises as to
what happens to all these VEVs once supersymmetry is broken.

We now sketch the fact that the supersymmetry breaking discussed
above, ensuing from $F$-terms, $F_S, F_T, F_{\phi_2} \neq 0$ and
driven by the non-perturbative superpotential, inevitably leads to a
stabilization of the many singlet `moduli' of the heterotic orbifold
vacuum. We shall consider here 3 classes of heterotic MSSM singlets.

\subsection{Singlets with polynomial Yukawa couplings}
\label{subsubsec:NONflats}

Let us first consider singlets having polynomial Yukawa couplings in
the superpotential, which in case of a coupling arising among purely
untwisted sector fields $\phi_i^{(U)}$ are perturbatively generated,
and in the other case involving {\it at least one} twisted sector
field $\phi_i^{(T)}$ are non-perturbatively generated (see Section
\ref{sec:modular}). The latter case is actually the most common
situation. Restricting again for reasons of simplicity to the case
of a single scalar field of the type under consideration, we can
describe the two cases as follows:

\begin{itemize}
  \item i) \[\fk=-3\log\left(T+\bar T-\bar\phi^{(U)}\phi^{(U)}\right)\quad,\quad \fw\supset \lambda\cdot\left(\phi^{(U)}\right)^N\;,\;N\geq3\]\\
  Note that the untwisted sector scalar fields $\phi^{(U)}$,
  being inherited from the bulk $\bf 248$ in 10d, appear this
  way in the K\"{a}hler potential.
  \item ii) \[\fk=-3\log(T+\bar T)+c\bar\phi^{(T)}\phi^{(T)}\quad,\quad \fw\supset e^{-b T}\left(\phi^{(T)}\right)^N\;,\;N\geq3\]\\
  Here the exponential dependence on $T$ arises from the
  $\eta$-function, which a non-perturbatively
  generated Yukawa coupling must have for reasons of modular
  invariance (see Section \ref{sec:modular}).
  \item iii) \[\fk=-3\log\left(T+\bar T-\bar\phi^{(U)}\phi^{(U)}\right)+c\,\bar\phi^{(T)}\phi^{(T)}\]\\[-4ex] \[\fw\supset \lambda e^{- b T}\left(\phi^{(T)}\right)^{N}+\tilde\lambda e^{- \tilde b T}\left(\phi^{(T)}\right)^{\tilde N}\left(\phi^{(U)}\right)^{M}\;{\rm with}\;\;M,N,\tilde N\geq2\]\\
  Here, too, the exponential dependence on $T$ from the
  $\eta$-function dependence of a non-perturbatively
  generated Yukawa coupling.
\end{itemize}

The calculation in case i) simplifies by the fact that there $K$
fulfills an extended no-scale relation
\bea \fk_{i}\fk^{i\bar j}\fk_{\bar j}&=&3\quad\forall\;i,j=T,\phi^{(U)}\nonumber\\
\fk^i=\fk^{i\bar j}\fk_{\bar j}&=&-\mathcal{V}\cdot\delta^i_T\quad,\quad
\mathcal{V}\equiv\left(T+\bar T-\bar\phi^{(U)}\phi^{(U)}\right) \eea
which implies for the F-term scalar potential a result \bea
V_F&=&e^\fk\left[\fk^{\phi^{(U)}\bar\phi^{(U)}}\left(|\partial_{\phi^{(U)}}\fw|^2+\left(\partial_{\phi^{(U)}}\fw\cdot\overline{\fk_{\phi^{(U)}}\fw}+c.c.\right)\right)\right.\nonumber\\
&&\qquad\left.+\frac{\mathcal{V}}{3}(T+\bar T)|\partial_T
\fw|^2+\left(\mathcal{V}\partial_T\fw+c.c.\right)\right]\quad. \eea
It is clear then that one solution to $\partial_{\phi^{(U)}}V_F=0$
is given by \be
\partial_{\phi^{(U)}}\fw=\partial_{\phi^{(U)}}\mathcal{V}=0\quad\Rightarrow\quad\langle\phi_{(U)}\rangle=0
\ee because $\partial_{\phi^{(U)}}\partial_{T}\fw\equiv 0\;\forall
\phi^{(U)}$. This implies that those untwisted sector singlets that
were stabilized at the origin in global supersymmetry by a purely
untwisted sector Yukawa coupling remain so even in supergravity.

For the twisted sector case ii) we find the scalar potential to be
\bea
V_F&=&e^\fk\left[\fk^{\phi^{(T)}\bar\phi^{(T)}}|D_{\phi^{(T)}}\fw|^2+\fk^{T\bar
T}\left(|\partial_T
\fw|^2+\underbrace{\partial_T\fw}_{\sim F_T}\overline{\fk_T\fw}+c.c.\right)\right]\nonumber\\
&\sim&e^{- 2 b T}\left(\bar\phi^{(T)}\phi^{(T)}\right)^{N-1}-
F_T(T+\bar T)e^{-b T}\left(\phi^{(T)}\right)^N+c.c. \eea which gives
two solutions to $\partial_{\phi^{(T)}}V_F=0$ as \be
\langle\phi^{(T)}\rangle=0\quad\bigvee\quad\langle\phi^{(T)}\rangle\sim\left(\frac{F_T
(T+\bar T)}{e^{-b
T}}\right)^{\frac{1}{N-2}}\sim\left(\frac{m_{3/2}}{e^{-b
T}}\right)^{\frac{1}{N-2}}\quad. \ee This implies that the
$\phi^{(T)}$ get stabilized either at the origin, or at non-zero but
small VEVs $\ll 1$. Their value in the latter case approaches
$\phi^{(T)}\sim M_{\rm GUT}$ for non-perturbative Yukawa couplings
of order $N\gtrsim 5$ and $m_{3/2}\sim {\rm TeV}$ (which can be
interesting for phenomenological reasons involving heavy vector-like
non-MSSM matter).

Finally, we note that case iii) reduces to case ii). To see this,
note, that the structure of \fk and $\fw$ given in case iii) does
not the change the arguments given for case i) which implies that in
case iii) we still find $\langle\phi_{(U)}\rangle=0$. This, however,
immediately gives us \be
\left.\fw\right|_{\langle\phi_{(U)}\rangle=0}\supset \lambda e^{-b T
}\left(\phi^{(T)}\right)^{N} \ee which is case ii).

\subsection{Singlet directions which are $F$- and $D$-flat in global supersymmetry}
\label{subsubsec:FDflats}

There are many directions in singlet field space in our heterotic
constructions which are $F$- and $D$-flat in global supersymmetry.
Let us denote these fields by - $\phi_i^{(f)}$, and the remaining
set of non-flat directions in field space by $\chi_i$. $D$-flatness
entails that the $D$-terms  do not depend on the $\phi_i^{(f)}$.
$F$-flatness implies that
$F_{\phi_i^{(f)}}=\partial_{\phi_i^{(f)}}\fw(\phi_i^{(f)},\chi_i)=const.$
for all values of $\langle \phi_i^{(f)} \rangle$.  Generically this
implies that $\langle \chi_i \rangle = 0$.

Simplifying to the case of a single $\chi$, this leads to a
consideration of  2 cases
\begin{itemize}
\item i)
\bea
F_{\phi_i^{(f)}}&=&0\quad\forall\, \phi_i^{(f)}\nonumber\\
\Rightarrow\;\fw  & \supset& e^{- b T} \ \chi \
\mathfrak{f}(\phi_i)\quad\bigvee\quad\fw  \supset  e^{- b T}
\chi^p \ \mathfrak{f}(\phi_i)\;,\; p \geq 2 \eea
\item ii)
\bea
F_{\phi_i^{(f)}}&=&const.\neq0\quad\forall\, \phi_i^{(f)}\nonumber\\
\Rightarrow\quad\fw&\supset& \lambda e^{-b  T} f(\tilde
\phi_j)\phi_i^{(f)} \eea
\end{itemize}
where the $\tilde \phi_j$ VEVs are assumed fixed by other terms in
the superpotential and $\mathfrak{f}$ is an arbitrary function of
its argument.

We consider first case i). At the supersymmetric minimum satisfying
$\partial_\chi \fw =
\partial_{\phi_i} \fw = 0$,  we have  $\langle \chi \rangle = 0$ with  $\langle
\phi_i \rangle$ arbitrary (subject, for the first case only, to the
condition $\mathfrak{f}(\phi_i) = 0$).  In this example we have $
\chi \in \{ \chi_i \} $ and $\phi_i \in \{ \phi_i^{(f)} \} $. Note
the fields $\phi_i^{(f)}$ effectively do not appear in the
superpotential at its minimum.

We now argue that the fields $\phi_i^{(f)}$ are stabilized by the
corrections from supergravity in the $F$-term scalar potential.
Namely, consider for sake of simplicity the case of a single such
field $\phi^{(f)}$ and $\chi$ with
\bea \fk&=&-3\log(T+\bar T)+ c\overline\phi^{(f)}\phi^{(f)} + c' \overline\chi \chi\nonumber\\
\partial_\chi \fw &=& \partial_{\phi^{(f)}} \fw \equiv 0\quad {\rm for} \quad \langle \chi \rangle =
0\eea we get the F-term scalar potential in supergravity to be (for
the twisted sector case ii) we find the scalar potential to be \bea
V_F&=&e^\fk\left(\fk^{\phi^{(f)}\bar\phi^{(f)}}|D_{\phi^{(f)}}\fw|^2+\fk^{\chi\bar\chi}|D_{\chi}\fw|^2
+\fk^{T\bar T}|D_T \fw|^2-3|\fw|^2\right)\nonumber\\
&=&e^\fk\left(c\overline\phi^{(f)}\phi^{(f)} - \kappa \right)\cdot|\fw|^2\nonumber\\
&\approx&|\fw|^2\cdot\left[-c(\kappa
-1)\overline\phi^{(f)}\phi^{(f)}-\frac{c^2(\kappa -
2)}{2}\left(\overline\phi^{(f)}\phi^{(f)}\right)^2+\right. \nonumber \\
&&\left.-\frac{c^3(\kappa-3)}{6}\left(\overline\phi^{(f)}\phi^{(f)}\right)^3
+\frac{c^4(4-\kappa)}{24}\left(\overline\phi^{(f)}\phi^{(f)}\right)^4+\ldots\right]
\eea Note, we maintain $\langle \chi \rangle = 0$, $\fw \neq 0$ is
due to other sectors of the theory and
$\kappa=(3-\fk^{T\bar T}|D_T \fw|^2/|\fw|^2)\leq3$ is a
positive semi-definite number of order 3. This scalar potential is
unbounded from above at large field values, $\phi^{(f)}$, thus
driving the VEV to large-field value. To this order in $V_F$ we find
\be \langle \phi^{(f)}\rangle\sim\frac{1}{\sqrt{c}}\quad. \ee This
implies that supergravity effects will serve to stabilize all the
globally supersymmetric {\it and} $F$- and $D$-flat singlet fields
generically at large values of ${\cal O}(1)$. Note, that the
non-perturbative effects coming from gaugino-condensation in the
hidden sector will add dependence of $\fw$ on $\phi^{(f)}$ beyond
the global mini-landscape analysis. This may render $\kappa$ a weak
function of $\phi^{f}$ such that we may for some of the globally
supersymmetric and $F$- and $D$-flat fields $\phi^{(f)}$ have
$\kappa<1$ at small $\phi^{(f)}$ while $1<\kappa<3$ at larger values
of $\phi^{(f)}$. In this situation the involved $\phi^{(f)}$-type
singlets will acquire vacua at both $\langle\phi^{(f)}\rangle=0$ and
$\langle\phi^{(f)}\rangle\sim1/\sqrt{c}$.  The $\chi$-like fields
will have their VEVs near the origin, i.e. they may be shifted from
the origin by small SUSY breaking effects.

Let us now turn to case ii) of $F$-flat but {\it non}-supersymmetric
singlet directions and look for vacua stabilizing $\phi^{(f)}\ll 1$
using again \be \fk=-3\log(T+\bar T)+ \overline\phi^{(f)}\phi^{(f)}
+ \overline\chi \chi\quad. \ee The scalar potential is \bea
V_F&=& e^\fk\left[\fk^{T\bar T}\underbrace{\left\langle D_{T}\fw\right\rangle}_{=F_T}\overline{\partial_T\fw}+c.c.+\fk^{\phi^{(f)}\bar\phi^{(f)}}|D_{\phi^{(f)}}\fw|^2\right]\nonumber\\
&\sim&\left\{\fk^{T\bar T}F_T\cdot b  \lambda e^{-b  T} f(\chi)\phi^{(f)}+c.c.\right.\nonumber\\
&&\left.+\fk^{\phi^{(f)}\bar\phi^{(f)}}\left[ \lambda e^{-b  T}
f(\chi)(1+\bar\phi^{(f)}\phi^{(f)})+\bar\phi^{(f)}\langle\fw\rangle\right]^2\right\}\quad.
\eea In the desired regime of $\phi^{(f)}\ll 1$ this gives us two
sub-cases:
\begin{itemize}
\item iia)
\[\fk^{\phi^{(f)}\bar\phi^{(f)}}F_{\phi^{(f)}}\ll \fk^{T\bar T}F_T\]
\item iib)
\[\fk^{\phi^{(f)}\bar\phi^{(f)}}F_{\phi^{(f)}}\gg \fk^{T\bar T}F_T\]
\end{itemize}
In case iia) $\phi^{(f)}\ll1$ implies that
$F_{\phi^{(f)}}\equiv\lambda e^{-b  T}
f(\langle\chi\rangle)\ll\langle\fw\rangle$ and thus
$\partial_{\phi^{(f)}}V_F=0$ gives us \be
\langle\phi^{(f)}\rangle\sim\frac{\langle
F_{\phi^{(f)}}\rangle}{\langle\fw\rangle}\ll1 \ee which is thus a
self-consistent vacuum.

In the opposite situation we get $F_{\phi^{(f)}}\equiv\lambda e^{-b
T} f(\langle\chi\rangle)\gg\langle\fw\rangle,\langle F_T\rangle$.
Using again $\phi^{(f)}\ll1$ this leads to \be
\langle\phi^{(f)}\rangle\sim\frac{\langle F_T\rangle}{\langle
F_{\phi^{(f)}}\rangle}\ll1\quad. \ee Thus, even the $F$-flat but
non-supersymmetric singlet directions of case ii) get stabilized by
supersymmetry breaking effects from the bulk moduli stabilization at
generically small but non-zero VEVs.

This property, of all $F$- and $D$-flat singlet fields generically
acquiring non-zero VEVs from supersymmetry breaking in the bulk
moduli stabilizing sector through supergravity, dynamically ensures
the decoupling of all vector-like non-MSSM matter at low-energies as
checked in global supersymmetry for the mini-landscape setup.

Note, that the overall vacuum structure of the F-flat singlet fields
implicates a choice of initial conditions. The amount of non-MSSM
vector-like extra matter in the mini-landscape constructions which
decouples from low energies depends on the choice of the globally
$F$-flat singlets $\phi^{(f)}_i$ placed at their non-zero VEV vacuum
instead of their zero VEV vacuum.  Thus, the choice of initial
conditions in the vacuum distribution among the set of globally
$F$-flat singlet fields characterizes how close to the MSSM one can
get when starting from one of the mini-landscape models.

Assuming now that one finds successful eternal inflation occurring
somewhere in the mini-landscape, this choice of initial conditions
turns into a question of cosmological dynamics. In this situation,
all possible initial conditions of the set of globally $F$-flat
singlets were potentially realized in a larger multiverse. The
choice of initial conditions on the singlets in the globally
$F$-flat sector would then be amenable to anthropic arguments and
might be eventually determined by selection effects.

%==============================================================
%==============================================================
\section{SUSY spectrum}
\label{sec:spectrum}
%==============================================================
%==============================================================

Now that we understand how SUSY is broken, we can calculate the
spectrum of soft masses.  The messenger of SUSY breaking is mostly
gravity, however, there are other contributions from gauge and
anomaly mediation.

\subsection{Contributions to the soft terms}

At tree level, the general soft terms for gravity mediation are
given in References
\cite{Brignole:1993dj,Brignole:1995fb,Kawamura:1996wn,Kawamura:1996bd,
Brignole:1997dp}. The models described in this paper contain an
additional contribution from the $F$-term of a scalar field
$\phi_2$.  Following References
\cite{Brignole:1993dj,Brignole:1995fb, Brignole:1997dp}, we define
\begin{equation} \label{F_terms_defn}
F^I \equiv e^{\fk/2}\fk^{I\bar{J}}\left(\bar{\fw}_{\bar{J}} + \bar{\fw}\fk_{\bar{J}}\right).
\end{equation}

\subsubsection{SUGRA effects}

{\em\textit{Gaugino masses}}

The tree level gaugino masses are given by
\begin{equation}
   M_a^{(0)} = \frac{g_a^2}{2}F^n\partial_n f_a(S) = \frac{g_a^2}{2}F^S.
\end{equation}
At tree level, the gauge kinetic function in heterotic string theory is linear
in the dilaton superfield $S$, and only dependent on the $T$ modulus at one loop.
It is important to note the enhancement of $F^S$ relative to $F_S$: naively, one
might guess that loop corrections to the gaugino masses might be important,
however
\begin{equation}
   F^S >> \frac{F^T}{16\pi^2},
\end{equation}
thus loop corrections will be neglected. \vspace{.1in}

\noindent{\em\textit{A Terms}}

At tree level, the $A$ terms are given by
\begin{equation}
   A_{IJK}^{(0)} = F^n \partial_n \fk + F^n\partial_n \log\frac{\fw_{IJK}}{\kappa_I \kappa_J \kappa_K},
\end{equation}
where
\begin{equation}
   \fw_{IJK} \equiv \frac{\partial^3 \fw}{\partial \Phi^I\partial\Phi^J\partial\Phi^K}
\end{equation}  and $\fk$ is the K\"{a}hler potential.  Neglecting $U$ dependence, we have
\begin{equation}
   \fk \supset \Phi_I\bar{\Phi}^I \prod_i \left(T_i+\bar{T}_i\right)^{-n^i_I} \Rightarrow \kappa_I \equiv \prod_i \left(T_i+\bar{T}_i\right)^{-n^i_I}.
\end{equation}  The $\kappa_I$ are the K\"{a}hler metrics for the chiral multiplets,
$\Phi_I$, where as the $A$ terms are expressed in terms of
canonically normalized fields.   As before, the modular weights of
the matter field are given by $n^i_I$.

In general, there are also tree level contributions to $A$ terms
proportional to \be - \frac{F_{\phi_2}}{\langle \phi_2 \rangle}
\frac{\partial \log \fw_{IJK}}{\partial \log \phi_2} . \ee These
terms may be dominant, but unfortunately they are highly model
dependent.  They may give a significant contribution to $A_b$ and
$A_\tau$, but in fact we find that the details of the low energy
spectrum are not significantly effected. \vspace{.1in}

\noindent{\em\textit{Scalar masses}}

The tree level scalar masses are given by
\begin{equation}
   \left(M_I^{(0)}\right)^2 = m_{3/2}^2 - F^n\bar{F}^{\bar{m}}\partial_n \partial_{\bar{m}} \log \kappa_I +
    g_G^2 \ f \ q^I_A \ \langle D_A \rangle \ \kappa_I  ,
\end{equation}
where $g_G^2 = 1/\re{S_0}$ and we have implicitly assumed that the
K\"{a}hler metric is diagonal in the matter fields. The factor $f$
re-scales the $U(1)_A$ charges $q_A$ from the mini-landscape
``benchmark" model 1 \cite{Lebedev:2007hv}, so they are consistent
with the charges $q_A'$ in our mini-version of the mini-landscape
model. We have $q_A' = q_A \ f = q_A \ \frac{48 \pi^2}{Tr Q}
\delta_{GS}$ with  $\delta_{GS} =\frac{N_f}{4 \pi^2}$ (Eqn.
\ref{eq:GS}) and $Tr Q = \frac{296}{3}$ (Eqn. E.5,
\cite{Lebedev:2007hv}) such that $\frac{Tr(q')}{4 \pi^2} =
\delta_{GS}$.

Again neglecting $U$ dependence, the K\"{a}hler metric for the
matter fields depends only on the $T$ moduli, and we find
\begin{equation} \label{tree_level_scalars}
   \left(M_I^{(0)}\right)^2 = m_{3/2}^2 - \sum_i \frac{n^i_I \left|F^{T_i}\right|^2}{\left(T_i+\bar{T}_i\right)^2} +
    g_G^2 \ f \ q^I_A \ \langle D_A \rangle / (2 \re{T_0})^{n^3_I} .
\end{equation}
\vspace{.1in}

\noindent{\em $\mu$ and $B\mu$ terms}

The $\mu$ term can come from two different sources:
\begin{equation}
\fk \supset Z(T_i+\bar{T}_i,U_j+\bar{U}_j,...) \mathbf{H}^u \mathbf{H}^d, \,\,\, \fw \supset
 \tilde \mu(\mathbf{s}_I,T_i,U_j,...) \mathbf{H}^u \mathbf{H}^d.
\end{equation}
In the orbifold models, K\"{a}hler corrections have not been
computed, so the function $Z$ is \textit{a priori} unknown.  Such a
term could contribute to the Giudice-Masiero mechanism
\cite{Giudice:1988yz}.  When both $\tilde \mu$ and $Z$ vanish, the
SUGRA contribution to the $\mu/B\mu$ terms vanish. On the other
hand, in the class of models which we consider, we know that vacuum
configurations exist such that $\tilde \mu = 0$ to a very high order
in singlet fields. Moreover $\tilde \mu \propto \langle \fw \rangle$
which vanishes in the supersymmetric limit, but obtains a value
$w_0$ at higher order in powers of chiral singlets.  If $\mu$ is
generated in this way, there is also likely to be a Peccei-Quinn
axion \cite{Kim:1983dt,Kim:1994eu}. Finally, supergravity effects
will also generate a $B \mu$ term. \vspace{.1in}

\noindent{\em Loop corrections}

Finally, one can consider loop corrections to the tree level
expressions in
\cite{Brignole:1993dj,Brignole:1995fb,Brignole:1997dp}.  This was
done in References \cite{Binetruy:2000md, Binetruy:2003ad}, where
the complete structure of the soft terms (at one loop) for a generic
(heterotic) string model were computed in the effective supergravity
limit. We have applied the results of \cite{Binetruy:2000md,
Binetruy:2003ad} to our models and find, at most, around a 10\%
correction to the tree level results of
\cite{Brignole:1993dj,Brignole:1995fb,Brignole:1997dp}.\footnote{In
estimating this result, we have assumed that the mass terms of the
Pauli-Villars fields do not depend on the SUSY breaking singlet
field $\phi_2$, and that the modular weights of the Pauli-Villars
fields obey specific properties.}

\subsubsection{Gauge mediation}

The ``mini-landscape" models generically contain vector-like exotics
in the spectrum.  Moreover it was shown that such states were
necessary for gauge coupling unification \cite{Dundee:2008ts}. The
vector-like exotics obtain mass in the supersymmetric limit by
coupling to scalar moduli, thus they may couple to the SUSY
breaking field $\phi_2$. We will consider the following light
exotics to have couplings linear in the field $\phi_2$:
\begin{equation} \label{light_exotica}
   n_3\times (\mathbf{3},1)_{1/3} + n_2 \times (1,\mathbf{2})_0 + n_1 \times (1,1)_{-1} + \text{h.c.}
\end{equation}
where the constants $n_i$ denote the multiplicity of states and (see
Table 7 of Reference \cite{Dundee:2008ts})
\begin{equation}
   n_3 \leq 4\text{~and~}n_2\leq 3\text{~and~}n_1 \leq 7.
\end{equation}

The gauge mediated contributions split the gaugino masses by an amount proportional
to the gauge coupling:
\begin{eqnarray}
   M_3^{(1)}|_{\text{gmsb}} &=& n_3 \frac{g_3^2}{16\pi^2}\frac{F^{\phi_2}}{\vev{\phi_2}}, \\
   M_2^{(1)}|_{\text{gmsb}} &=& n_2 \frac{g_2^2}{16\pi^2}\frac{F^{\phi_2}}{\vev{\phi_2}}, \\
   M_1^{(1)}|_{\text{gmsb}} &=& \frac{n_3+3n_1}{10} \frac{g_1^2}{16\pi^2}\frac{F^{\phi_2}}{\vev{\phi_2}}.
\end{eqnarray}
It is interesting to note that this becomes more important as
$\vev{\phi_2}$ decreases/$F^{\phi_2}$ increases, or if there are a
large number of exotics present.

The scalar masses in gauge mediation come in at two loops, and receive corrections
proportional to
\begin{equation}
   \left(M_I\right)^2|_{\text{gmsb}} \sim \left(\frac{1}{16\pi^2}\right)^2\left(\frac{F_{\phi_2}}{\phi_2}\right)^2.
\end{equation}
Unlike in the case of the gaugino masses, however, the tree level
scalar masses are set by the gravitino mass.  Typically
\begin{equation}
   16\pi^2 m_{3/2} >> \frac{F_{\phi_2}}{\phi_2},
\end{equation}
and the gauge mediation contribution gives about a 10\% correction to the scalar masses,
in our case.  We will neglect their contributions in the calculation of the soft masses below.

\subsection{Calculation of the soft terms -- relevant details from the ``mini-landscape"}
\label{sec:details}

 Given the relative sizes of the $F$-terms in the
SUSY breaking sectors described in this paper, it is very difficult
to make model-independent statements.  This stems from the fact that
$F^T$ plays a dominant role in the SUSY breaking.  Because the
K\"{a}hler metrics for the matter fields have generally different
dependences on the $T$ modulus, the dependence of the soft terms on
$F^T$ is typically non-universal.  Moreover, the couplings of the
SUSY breaking singlet field $\phi_2$ will necessarily depend on the
details of a specific model.  Thus, in order to make \textit{any}
statements about the phenomenology of these models, we will have to
make some assumptions.  With the general features of the
``mini-landscape'' models in mind, we will make the following
assumptions:
\begin{enumerate}
   \item SUSY breaking is dominated by $F_{\phi_2} \neq 0$, $F_{T_3} \neq 0$, $F_S\neq 0$.
All other $F$ terms, including those due to the other $T$ and $U$ moduli, are subdominant;
   \item \label{brane_exotics_oinly} the massless spectrum below $\mstring$ contains some vector-like
exotics;
   \item the untwisted sector contains the following Higgs and (3rd generation) matter multiplets:
$\mathbf{H}_u, \mathbf{H}_d, \mathbf{Q}_3, \mathbf{U}^c_3,
\mathbf{E}^c_3$;
   \item the first two families have the same modular weights, see Table
   \ref{tab:modular_weights_MSSM};
   \item the SUSY breaking field, $\phi_2$, lives in the untwisted, or second or fourth twisted sector, with a modular weight given by $n^3 = 0$; and
\item we neglect possible $\phi_2$ dependence of the effective
Yukawa terms.
\end{enumerate}
Let us examine these assumptions in some more detail.

In general, gauge coupling unification in the ``mini-landscape"
models seems to require the existence of light vector-like exotics
\cite{Dundee:2008ts}, whose masses can be as small as
$\order{10^{9}\gev}$.  We further assume that these exotics couple
to the SUSY breaking field $\phi_2$, giving a gauge mediated
contribution to the gaugino masses above.  We will make this
contribution to the soft terms explicit in what follows. In
assumption \ref{brane_exotics_oinly} we have specialized to the case
where only ``brane-localized'' exotics are present in the model.
These are states which come from the first and third twisted sectors
of the model, and we refer the reader to \cite{Lebedev:2007hv,
Dundee:2008ts} for more details.

The top quarks and the up Higgses live in the bulk and the string
selection rules allow for the following coupling in the
superpotential:
\begin{equation} \label{third_family_top}
   \fw \supset c\, \mathbf{Q}_3 \mathbf{H}_u \mathbf{U}^c_3.
\end{equation}
The coupling $c$ is a pure number of \order{1}, and is free of any
dependence on the moduli.  The down and lepton Yukawas are a bit
more involved, as they arise at a higher order in the stringy
superpotential. We will take them to be of the following form:
\begin{equation} \label{third_family_bottom}
   \fw \supset \eta(T_1)^{p_1}\eta(T_2)^{p_2}\eta(T_3)^{p_3} \
   \left(f_1(\vev{\mathbf{s}_I^5})\mathbf{Q}_3 \mathbf{H}_d \mathbf{D}^c_3 + f_2(\vev{\mathbf{s}_I^5})\mathbf{L}_3 \mathbf{H}_d \mathbf{E}^c_3\right).
\end{equation}
The $\mathbf{s}_I$ are other singlet fields in the model (excluding
the SUSY breaking singlet field, $\phi_2$, as per our assumptions),
and the numbers $p_1,p_2$ and $p_3$ are calculable in principle,
given knowledge of the modular weights of the $\mathbf{s}_I$. As one
might expect, the expressions for the $A$ terms explicitly depend on
the value of $p_3$ in such a way that changing its value may result
in a significant change in $A_b$ and $A_{\tau}$ at the string scale.
The impact on the weak scale observables is much less severe,
however, giving a correction of a few percent to the gaugino masses,
and leaving the squark and slepton masses virtually unchanged.
Motivated by the modular weight assignments in Table
\ref{tab:modular_weights_MSSM}, we will choose $p_3 = 0$.  Note this
choice gives us universal $A$ terms for the third generation.

One of the nice features of the ``mini-landscape" models is the
incorporation of a discrete ($D_4$) symmetry between the first two
families in the low energy effective field theory.  Because of this
symmetry, we expect the modular weights of these matter states to be
the same \cite{Ko:2007dz}, see Table \ref{tab:modular_weights_MSSM}.
This will turn out to be very beneficial in alleviating the flavor
problems that are generic in gravity mediated models of SUSY
breaking: the scalar masses (at tree level) are given by a universal
contribution (the gravitino mass squared) plus a contribution
proportional to the modular weight.  If the modular weights are the
same between the first two generations, then the leading order
prediction is for degenerate squark and slepton masses in the two
light generations. Other contributions to the scalar masses come
from gauge mediation and anomaly mediation, which do not introduce
any new flavor problems into the low energy physics.

\begin{table}
\centering
   \begin{tabular}{|c|c|cc|}
\hline
MSSM particle&Modular Weight $\vec{n}$& \multicolumn{2}{c|}{$U(1)_A$ charge}\\
\hline  $\mathbf{Q}_3$&$(0,1,0)$ & \multicolumn{2}{c|}{$4/3$} \\
$\mathbf{U}^c_3$&$(1,0,0)$ & \multicolumn{2}{c|}{$2/3$} \\
$\mathbf{D}^c_3$&$\left(\frac{1}{3},\frac{2}{3},0\right)$ & \multicolumn{2}{c|}{$8/9$} \\
$\mathbf{L}_3$&$\left(\frac{2}{3},\frac{1}{3},0\right)$ & \multicolumn{2}{c|}{$4/9$} \\
$\mathbf{E}^c_3$ & $(1,0,0)$ & \multicolumn{2}{c|}{$2/3$} \\
\multirow{2}{*}{first two gen.}&\multirow{2}{*}{$\left(\frac{5}{6},\frac{2}{3},\frac{1}{2}\right)$}
& $7/18$ &  (${\bf 10}$) \\
 &&  $-5/18$ & (${\bf \bar 5}$) \\
 $\mathbf{H}_u$&$(0,0,1)$ & \multicolumn{2}{c|}{$-2$} \\
 $\mathbf{H}_d$&$(0,0,1)$ & \multicolumn{2}{c|}{$+2$} \\
 \hline \hline \end{tabular}
\caption{\label{tab:modular_weights_MSSM}Modular weights of the MSSM
states in the ``mini-landscape" benchmark model 1A.  For the first
two generations, the $U(1)_A$ charges differ depending on whether
the particle is in the {\bf 10} or ${\bf \bar 5}$ of $SU(5)$. See
\cite{Lebedev:2007hv} for details.} \end{table}

\subsection{Hierarchy of $F$-terms}

Note, in Section \ref{sec:one_gaugino_condensate}, we find (roughly)
\begin{equation}
 F_T >>  F_S \gtrsim F_{\phi_2},
\end{equation}  for Cases 1, 2 and 3; and
\begin{equation}
 F_T \gtrsim F_{\phi_2} >> F_S,
\end{equation}  for Cases 4 and 5, where
\begin{equation}
F_I \equiv \fw_I + \fw \fk_I.
\end{equation}
When one includes the relevant factors of the K\"{a}hler metric, we
have (Table \ref{tab:F_terms})
\begin{equation}
F^T > F^S >> F^{\phi_2}
\end{equation} for Cases 1, 2 and 3; and
\begin{equation}
F^T >> F^S \sim F^{\phi_2}
\end{equation} for Cases 4 and 5.
$F^S$ is enhanced by a factor of $\fk^{S\bar{S}}\sim (2+2)^2$, while
$F^{\phi_2}$ is decreased by a factor of
$\fk^{\phi_2\bar{\phi}_2}\sim(2)^{-1/2}$.\footnote{This is due to
the assumed modular weight of the field $\phi_2$ (assumption 5 in
Section \ref{sec:details}).} This means that although the singlet
field $\phi_2$ was a dominant source of SUSY breaking, it is the
least important when computing the soft terms, given the one
condensate hidden sector of the known ``mini-landscape" models
studied in Section \ref{sec:one_gaugino_condensate}.\footnote{In
racetrack models $F_S$ is suppressed by more than an order of
magnitude. In these cases $F_{\phi_2}$ is dominant
\cite{Gallego:2008sv}.}   Taking the details of the
``mini-landscape" models into account, the soft terms at the string
scale are given in Table \ref{tab:string_scale_BCs}.

%================================================================
\begin{table}[t!]
\renewcommand\arraystretch{1.2}
\centering
\noindent\makebox[\textwidth]{ \scalebox{0.75}{
\begin{tabular}{|c|c|c|c|c|c|}
\hline
 &  Case 1 &  Case 2 & Case 3 & Case 4 & Case 5 \\
\hline
$F^S$  & $6.6\times10^{-16}$ & $3.7\times10^{-16}$ & $4.2\times10^{-16}$ & $2.7\times10^{-16}$ & $2.1\times10^{-16}$\\
\hline
$F^T$  & $-2.2\times10^{-15}$ & $-1.2\times10^{-15}$ & $-1.4\times10^{-15}$ & $1.6\times10^{-15}$ & $2.2\times10^{-15}$\\
\hline
$F^{\phi_2}$ & $-1.1\times10^{-17}$ & $-6.5\times10^{-18}$ & $-7.7\times10^{-18}$ & $1.9\times10^{-16}$ & $1.8\times10^{-16}$\\
\hline
\end{tabular}
}} \caption{\label{tab:F_terms} The hierarchy of $F$ terms in the
five examples of the single condensate model we studied.  Note that
$F^{\Phi}$ is defined in Eqn. (\ref{F_terms_defn}).  All of the $F$
terms contribute to the soft masses, as they are all within an order
of magnitude.}
\end{table}
%===================================================================

In the five chosen Cases, 2, 3 and 4 have a gravitino mass less than
2 TeV.  The value of the gravitino mass can be adjusted by varying
$w_0$.  For Cases, 1, 3 (4) the Higgs up (down) mass squared is
negative. This is a direct result of the sign of $D_A$ and the
$U(1)_A$ charge of the Higgs' (see Table
\ref{tab:modular_weights_MSSM} for the $U(1)_A$ charges of all the
MSSM states).\footnote{Note, it is well known that the $D$-term VEV
in supergravity is of order $\langle F^i \rangle^2$
\cite{Kawamura:1995ec,Kawamura:1996bd}. It is given by the relation
\be \langle D_A \rangle = 2 M_A^{-2} \langle F^i \rangle \langle
F_j^* \rangle \langle \partial_i \partial^j D_A \rangle . \ee Thus
the $D$-term contribution to the vacuum energy is negligible, but
its contribution to scalar masses can be significant. Since $|F^S|^2
< |F^T|^2$, $F^T$ is dominant in the above relation. However, the
K\"{a}hler metric of $\phi_2$ which spontaneously breaks $U(1)_A$,
in our case, does not include $T$, i.e. $ \langle (\partial_T
\partial^T D_A) \rangle = 0$.  Hence $\langle D_A \rangle$ is
suppressed compared with $|F^T|^2/M_{Pl}^2$, i.e. $\langle D_A
\rangle : |F^T|^2/M_{Pl}^2 = |F^S|^2 : |F^T|^2$ where we used
$\langle (\partial_S \partial^S D_A)\rangle = (M_A/M_{Pl})^2$,
because of the S-dependent FI term.  We thank T.~Kobayashi, private
communication, for this analysis.  However, it should be clear that
we have also used the freedom available in the Coleman-Weinberg
one-loop correction to further adjust the value of the $D$-term.}
Note, the first and second generation squarks and sleptons are
lighter than the third generation states at the string scale. This
is a consequence of the significant $T$ modulus contribution to the
first and second generation squark and slepton masses, due to their
modular weights, Table \ref{tab:modular_weights_MSSM}. Finally we
have included the possible gauge mediated SUSY breaking contribution
to the gaugino masses, Table \ref{tab:string_scale_BCs}.  This
contribution is only significant for Cases 4 and 5, due to the
larger value of $F_{\phi_2}$ in these cases.

\begin{table}[t!]
\renewcommand\arraystretch{1.2}
\centering \noindent\makebox[\textwidth]{ \scalebox{0.75}{
   \begin{tabular}{|c|cc|cc|cc|cc|cc|}
\cline{2-11}
\multicolumn{1}{c|}{ }&\multicolumn{10}{c|}{\textit{All Masses in GeV}} \\
\hline
Parameter&\multicolumn{2}{c|}{Case 1}&\multicolumn{2}{c|}{Case 2}&\multicolumn{2}{c|}{Case 3}&\multicolumn{2}{c|}{Case 4}&\multicolumn{2}{c|}{Case 5}\\
\hline
$m_{3/2}$&\multicolumn{2}{c|}{2159}&\multicolumn{2}{c|}{1350}&\multicolumn{2}{c|}{1133}&\multicolumn{2}{c|}{1808}&\multicolumn{2}{c|}{2375}\\
\hline
$m_{H_u}$&\multicolumn{2}{c|}{$478i$}&\multicolumn{2}{c|}{168}&\multicolumn{2}{c|}{$372i$}&\multicolumn{2}{c|}{688}&\multicolumn{2}{c|}{384}\\
$m_{H_d}$&\multicolumn{2}{c|}{679}&\multicolumn{2}{c|}{216}&\multicolumn{2}{c|}{495}&\multicolumn{2}{c|}{$476i$}&\multicolumn{2}{c|}{251}\\
\hline
$M_1$&\multicolumn{2}{c|}{$362-0.3n_1-0.1n_3$}&\multicolumn{2}{c|}{$206-0.2n_1-0.1n_3$}&\multicolumn{2}{c|}{$243-0.2n_1-0.1n_3$}&\multicolumn{2}{c|}{$158+13n_1+4n_3$}&\multicolumn{2}{c|}{$118+7n_1+2n_3$}\\
$M_2$&\multicolumn{2}{c|}{$362-1n_2$}&\multicolumn{2}{c|}{$206+1n_2$}&\multicolumn{2}{c|}{$243-1n_2$}&\multicolumn{2}{c|}{$158+45n_2$}&\multicolumn{2}{c|}{$118+23n_2$}\\
$M_3$&\multicolumn{2}{c|}{$362-1n_3$}&\multicolumn{2}{c|}{$206+1n_3$}&\multicolumn{2}{c|}{$243-1n_3$}&\multicolumn{2}{c|}{$158+45n_3$}&\multicolumn{2}{c|}{$118+23n_3$}\\
\hline
$A_t$&\multicolumn{2}{c|}{3901}&\multicolumn{2}{c|}{2466}&\multicolumn{2}{c|}{1974}&\multicolumn{2}{c|}{-3690}&\multicolumn{2}{c|}{-4798}\\
$A_{b}$&\multicolumn{2}{c|}{3901}&\multicolumn{2}{c|}{2466}&\multicolumn{2}{c|}{1974}&\multicolumn{2}{c|}{-3690}&\multicolumn{2}{c|}{-4798}\\
$A_{\tau}$&\multicolumn{2}{c|}{3901}&\multicolumn{2}{c|}{2466}&\multicolumn{2}{c|}{1974}&\multicolumn{2}{c|}{-3690}&\multicolumn{2}{c|}{-4798}\\
\hline
&Gen.~1,2&Gen.~3&Gen.~1,2&Gen.~3&Gen.~1,2&Gen.~3&Gen.~1,2&Gen.~3&Gen.~1,2&Gen.~3\\
\cline{2-11}
$m_{\tilde{q}}$     & 1580 & 2288 &  966 & 1355 &  895 & 1446 & 1262 & 1657 & 1691 & 2361 \\
$m_{\tilde{u}^c}$   & 1580 & 2225 &  966 & 1353 &  895 & 1299 & 1262 & 1734 & 1691 & 2368 \\
$m_{\tilde{d}^c}$   & 1521 & 2246 &  964 & 1354 &  757 & 1350 & 1330 & 1709 & 1697 & 2366 \\
$m_{\tilde{\ell}}$  & 1580 & 2203 &  964 & 1352 &  757 & 1246 & 1330 & 1759 & 1697 & 2370 \\
$m_{\tilde{e}^c}$   & 1580 & 2225 &  966 & 1353 &  895 & 1299 & 1262 & 1734 & 1691 & 2368 \\
\hline
\hline
\end{tabular}
} } \caption{\label{tab:string_scale_BCs}Boundary conditions at the
string scale. $n_3,n_2,n_1$ refer to possible intermediate mass
vector-like exotics which couple to the SUSY breaking field
$\phi_2$, see Eqn. (\ref{light_exotica}).}
\end{table}

\subsection{Weak scale observables}

We do not intend this work to be a comprehensive study of the
parameter space of these models, so we will limit our weak scale
analysis to the five cases studied in the single condensate model
presented in this paper.  The points are chosen subject to the
following constraints:
\begin{itemize}
   \item $m_{h^0}\Big|_{\text{LEP}} \gtrsim 114.4 \gev$,
   \item successful electroweak symmetry breaking,
   \item $m_{\tilde{\chi}^{\pm}} \gtrsim 94\gev$, and
   \item the low energy spectrum is free of tachyons.
\end{itemize}
Note that we take $\text{sgn}(\mu) > 0$ and vary $\tan\beta$, and
the number, $n_i$, of ``messenger'' exotics.  We stay in the region
of small to moderate $\tan\beta$ as the ``mini-landscape" models do
not tend to predict unification of the third family Yukawas.  This
can be seen from Eqns. (\ref{third_family_top}) and
(\ref{third_family_bottom}), for example.

Using {\tt SoftSUSY} (v3.1) \cite{Allanach:2001kg}, we preformed the RGE running
from the string scale to the weak scale.  We use the current value of the top quark
mass \cite{tev_EW_working_group}
\begin{equation}
   m_{top}\Big|_{\text{world~avg.}} = 173.1 \gev
\end{equation}
and the strong coupling constant at $\mz$ \cite{Amsler:2008zzb}
\begin{equation}
   \alpha_s(\mz) = 0.1176.
\end{equation}
The $\mu$ parameter is obtained under the requirement of radiative electroweak
symmetry breaking, and is of order the gravitino mass, as expected.  This implies
a fine tuning of order
\begin{equation}
   \frac{\mz^2}{m_{3/2}^2}\sim \order{10^{-2}} \text{~to~} \order{10^{-4}}.
\end{equation}

The results obtained from \texttt{SoftSUSY} are presented in Table
\ref{tab:weak_obs}.   In this analysis, we have not included any
possible gauge mediated SUSY breaking contributions.  This assumes
that all the vector-like exotics have mass at the string scale.  In
Case 2 and 3 we have the smallest gravitino masses, so the lightest
SUSY partners.  $\tan\beta = 25$ in order for the light Higgs mass
to be above the LEP bound.  Note we assume a $\pm 2$ GeV theoretical
uncertainty in the Higgs mass.  In all 5 cases the Higgs mass is
between the LEP bound and 121 GeV.  All other Higgs masses are of
order the gravitino mass.   In all 5 cases the gluino mass is less
than 1 TeV and of order 600 GeV or less in Cases 2, .. , 5.  Thus
the gluino is very observable at the LHC.  In all cases, the
lightest MSSM particle is the lightest neutralino. The
next-to-lightest neutralino and the lightest chargino are
approximately degenerate with mass of order twice the lightest
neutralino mass.  In Cases 2, 3 and 4 the lightest stop has mass
less than 1 TeV.   In Cases 2 and 4, the lightest stop is also the
lightest squark.   Thus in these cases the gluino will predominantly
decay into a top - anti-top pair with missing energy (and possibly
two energetic leptons).    In Case 3, the lightest down squarks of
the first two families are lighter than the lightest stop.   In
these cases gluinos will decay significantly into two light quark
jets plus missing energy (and possibly two energetic leptons).

%================================================================
\begin{table}[t!]
\renewcommand\arraystretch{1.2}
\begin{center}
\noindent\makebox[\textwidth]{ \scalebox{0.75}{
   \begin{tabular}{c|c|c|c|c|c|c|c|c|c|c|c|}
\cline{3-12}
\multicolumn{2}{c|}{ }&\multicolumn{10}{c|}{\textit{All Masses in GeV  (defined at  $M_{\textsc{w}} \approx 80$ GeV, unless otherwise noted.) }} \\
\hline
&Observable&\multicolumn{2}{c|}{Case 1}&\multicolumn{2}{c|}{Case 2}&\multicolumn{2}{c|}{Case 3}&\multicolumn{2}{c|}{Case 4}&\multicolumn{2}{c|}{Case 5}\\
\hline
\multirow{4}{*}{\rotatebox{90}{Inputs}}&$m_{3/2}$&\multicolumn{2}{c|}{2159}&\multicolumn{2}{c|}{1350}&\multicolumn{2}{c|}{1133}&\multicolumn{2}{c|}{1808}&\multicolumn{2}{c|}{2375}\\
&$\tan\beta$&\multicolumn{2}{c|}{10}&\multicolumn{2}{c|}{25}&\multicolumn{2}{c|}{25}&\multicolumn{2}{c|}{10}&\multicolumn{2}{c|}{4}\\
&$\text{sgn}(\mu)$&\multicolumn{2}{c|}{$+$}&\multicolumn{2}{c|}{$+$}&\multicolumn{2}{c|}{$+$}&\multicolumn{2}{c|}{$+$}&\multicolumn{2}{c|}{$+$}\\
&$n_1,n_2,n_3$&\multicolumn{2}{c|}{0,0,0}&\multicolumn{2}{c|}{0,0,0}&\multicolumn{2}{c|}{0,0,0}&\multicolumn{2}{c|}{0,0,0}&\multicolumn{2}{c|}{0,0,0}\\
\hline
\multirow{5}{*}{\rotatebox{90}{EWSB}}&$\mu(\msusy)$&\multicolumn{2}{c|}{2221}&\multicolumn{2}{c|}{1317}&\multicolumn{2}{c|}{1342}&\multicolumn{2}{c|}{1848}&\multicolumn{2}{c|}{2636}\\
&$m_{h^0}$&\multicolumn{2}{c|}{115.8}&\multicolumn{2}{c|}{113.3}&\multicolumn{2}{c|}{113.5}&\multicolumn{2}{c|}{121.4}&\multicolumn{2}{c|}{116.7}\\
&$m_{H^0}$&\multicolumn{2}{c|}{2299}&\multicolumn{2}{c|}{1161}&\multicolumn{2}{c|}{1368}&\multicolumn{2}{c|}{1731}&\multicolumn{2}{c|}{2717}\\
&$m_{A^0}$&\multicolumn{2}{c|}{2303}&\multicolumn{2}{c|}{1173}&\multicolumn{2}{c|}{1376}&\multicolumn{2}{c|}{1728}&\multicolumn{2}{c|}{2715}\\
&$m_{H^+}$&\multicolumn{2}{c|}{2305}&\multicolumn{2}{c|}{1176}&\multicolumn{2}{c|}{1379}&\multicolumn{2}{c|}{1730}&\multicolumn{2}{c|}{2716}\\
\hline
\multirow{3}{*}{\rotatebox{90}{$M_a(\msusy)$}} &$M_1$&\multicolumn{2}{c|}{151}&\multicolumn{2}{c|}{83}&\multicolumn{2}{c|}{100}&\multicolumn{2}{c|}{68}&\multicolumn{2}{c|}{53}\\
&$M_2$&\multicolumn{2}{c|}{277}&\multicolumn{2}{c|}{155}&\multicolumn{2}{c|}{185}&\multicolumn{2}{c|}{128}&\multicolumn{2}{c|}{100}\\
&$M_3$&\multicolumn{2}{c|}{773}&\multicolumn{2}{c|}{457}&\multicolumn{2}{c|}{538}&\multicolumn{2}{c|}{370}&\multicolumn{2}{c|}{279}\\
\hline
\rotatebox{90}{$\tilde{g}$}&$m_{\tilde{g}}$&\multicolumn{2}{c|}{914}&\multicolumn{2}{c|}{545}&\multicolumn{2}{c|}{630}&\multicolumn{2}{c|}{456}&\multicolumn{2}{c|}{365}\\
\hline
\multirow{6}{*}{\rotatebox{90}{Neut./Charg.}}&$m_{\tilde{\chi}_1^0}$&\multicolumn{2}{c|}{150}&\multicolumn{2}{c|}{83}&\multicolumn{2}{c|}{99}&\multicolumn{2}{c|}{68}&\multicolumn{2}{c|}{52}\\
&$m_{\tilde{\chi}_2^0}$&\multicolumn{2}{c|}{293}&\multicolumn{2}{c|}{164}&\multicolumn{2}{c|}{194}&\multicolumn{2}{c|}{136}&\multicolumn{2}{c|}{104}\\
&$m_{\tilde{\chi}_3^0}$&\multicolumn{2}{c|}{-2204}&\multicolumn{2}{c|}{-1306}&\multicolumn{2}{c|}{-1334}&\multicolumn{2}{c|}{-1835}&\multicolumn{2}{c|}{-2616}\\
&$m_{\tilde{\chi}_4^0}$&\multicolumn{2}{c|}{2206}&\multicolumn{2}{c|}{1307}&\multicolumn{2}{c|}{1335}&\multicolumn{2}{c|}{1836}&\multicolumn{2}{c|}{2617}\\ \cline{2-12}
&$m_{\tilde{\chi}_1^\pm}$&\multicolumn{2}{c|}{293}&\multicolumn{2}{c|}{164}&\multicolumn{2}{c|}{194}&\multicolumn{2}{c|}{136}&\multicolumn{2}{c|}{104}\\
&$m_{\tilde{\chi}_2^\pm}$&\multicolumn{2}{c|}{2214}&\multicolumn{2}{c|}{1313}&\multicolumn{2}{c|}{1341}&\multicolumn{2}{c|}{1839}&\multicolumn{2}{c|}{2622}\\
\hline
\multirow{8}{*}{\rotatebox{90}{Squarks/Sleptons}}&& Gen.~1,2&Gen.~3& Gen.~1,2&Gen.~3& Gen.~1,2&Gen.~3& Gen.~1,2&Gen.~3& Gen.~1,2&Gen.~3\\ \cline{3-12}
&$m_{\tilde{u}_1}$  &  1712  &  1542  &  1040  &   921  &  1013  &   987  &  1283 &   717 &  1677  &  1107  \\
&$m_{\tilde{u}_2}$  &  1704  &  2042  &  1038  &  1164  &  1006  &  1336  &  1289 &  1260 &  1683  &  1860  \\
&$m_{\tilde{d}_1}$  &  1714  &  2037  &  1043  &  1150  &  1016  &  1316  &  1285 &  1223 &  1678  &  1838  \\
&$m_{\tilde{d}_2}$  &  1651  &  2321  &  1036  &  1341  &   888  &  1379  &  1351 &  1702 &  1688  &  2364  \\
&$m_{\tilde{e}_1}$  &  1532  &  2192  &   970  &  1227  &   769  &  1182  &  1334 &  1696 &  1694  &  2356  \\
&$m_{\tilde{e}_2}$  &  1586  &  2206  &   968  &  1305  &   901  &  1228  &  1256 &  1750 &  1687  &  2370  \\
&$m_{\tilde{\nu}}$  &  1530  &  2196  &   966  &  1296  &   764  &  1202  &  1331 &  1746 &  1692  &  2366  \\
\hline
\multirow{6}{*}{\rotatebox{90}{Other Obs.}}&
$\delta\rho$&\multicolumn{2}{c|}{$8.5\times10^{-6}$}&\multicolumn{2}{c|}{$3.0\times10^{-5}$}&\multicolumn{2}{c|}{$2.3\times10^{-5}$}&\multicolumn{2}{c|}{$2.1\times10^{-5}$}&\multicolumn{2}{c|}{$7.2\times10^{-6}$}\\
&$\delta(g-2)_{\mu}$&\multicolumn{2}{c|}{$6.0\times10^{-11}$}&\multicolumn{2}{c|}{$3.9\times10^{-10}$}&\multicolumn{2}{c|}{$5.5\times10^{-10}$}&\multicolumn{2}{c|}{$7.0\times10^{-11}$}&\multicolumn{2}{c|}{$1.2\times10^{-11}$}\\
&$BR(b\rightarrow s\gamma)$&\multicolumn{2}{c|}{$3.7\times10^{-4}$}&\multicolumn{2}{c|}{$3.9\times10^{-4}$}&\multicolumn{2}{c|}{$3.9\times10^{-4}$}&\multicolumn{2}{c|}{$3.6\times10^{-4}$}&\multicolumn{2}{c|}{$3.7\times10^{-4}$}\\
&$BR(B_s \rightarrow \mu^+\mu^-)$&\multicolumn{2}{c|}{$3.1\times10^{-9}$}&\multicolumn{2}{c|}{$2.7\times10^{-9}$}&\multicolumn{2}{c|}{$2.9\times10^{-9}$}&\multicolumn{2}{c|}{$3.1\times10^{-9}$ }&\multicolumn{2}{c|}{$3.1\times10^{-9}$}\\ \cline{2-12}
&$m_{LMM}$&\multicolumn{2}{c|}{272}&\multicolumn{2}{c|}{175}&\multicolumn{2}{c|}{138}&\multicolumn{2}{c|}{531}&\multicolumn{2}{c|}{487}\\
&$m_{nLMM}$&\multicolumn{2}{c|}{41659}&\multicolumn{2}{c|}{25694}&\multicolumn{2}{c|}{22745}&\multicolumn{2}{c|}{27231}&\multicolumn{2}{c|}{36795}\\
\hline
\hline
\end{tabular}
} }
\end{center}
\caption{\label{tab:weak_obs} \small Weak scale observables, with no
contribution from gauge mediation: $n_3 = n_2 = n_1 = 0$ , see Eqn.
\ref{light_exotica}.  We have listed the mass eigenstates of the
squarks and sleptons.  Note that for light generations,
$m_{\tilde{u}_1} \approx m_{\tilde{u}_L}$, etc. The last two rows
give the lightest massive modulus ($m_{LMM}$)[mostly K\"{a}hler
modulus (\re{T})]  and the \textit{next to} lightest massive modulus
($m_{nLMM}$) [mostly the dilaton (\re{S})]. All other moduli have
mass $\gtrsim 100$ TeV.}
\end{table}
%================================================================

\FloatBarrier

In all cases the lightest MSSM particle is mostly ($\gtrsim 99\%$)
bino (see Table \ref{tab:param_space}). We note that this is
generically true in the models, even when there are contributions
from gauge mediation.  The gauge mediated contributions in Eqn.
(\ref{light_exotica}) do not appreciably change the composition of
the LSP, which one can check with the solutions in Table 7 of
Reference \cite{Dundee:2008ts}.

We have evaluated other low energy observables using {\tt
micrOMEGAs} \cite{Belanger:2008sj}. As expected, the bino LSP
overcloses the universe, giving $\Omega_{\textsc{dm}}
>> \Omega_{\textsc{dm}}^{\textsc{obs}}\approx 0.2$. The calculated values
for the following observables are given in the last few rows of
Table \ref{tab:weak_obs}.  Corrections to the $\rho$ parameter are
very small.  Corrections to $(g-2)_{\mu}$ are significant in Cases 2
and 3 which is not surprising since these are the two cases with the
lightest sleptons for the first two families.  We also display the
results for $BR(b\rightarrow s\gamma)$ and $BR(B_s \rightarrow
\mu^+\mu^-)$. The result for $BR(b\rightarrow s\gamma)$ is within
the 2$\sigma$ experimental bound (see \cite{Misiak:2009nr} and
references therein). Given the small chargino masses and the large
values of $\mu$ and the squark and CP odd Higgs masses, we obtain a
branching ratio $BR(B_s \rightarrow \mu^+ \mu^-)$ consistent with
the standard model.

We are not overly concerned about the fact that binos seem to
overclose the universe.  In some of the heterotic orbifold models
the Higgs $\mu$ term vanishes in the supersymmetric limit. Hence
there is a Peccei-Quinn symmetry. Supersymmetry breaking effects are
expected to shift the moduli VEVs and generate a non-vanishing $\mu$
term; spontaneously breaking the PQ symmetry and producing the
standard invisible axion.  In fact, it has been shown that PQ axions
may be obtained in heterotic orbifold contructions
\cite{Choi:2009jt}. In such cases it is possible that the bino
decays to an axino + photon leaving an axino dark matter candidate
\cite{Rajagopal:1990yx,Covi:1999ty,Covi:2001nw}.

However another, perhaps more important, cosmological effect must be
considered.   All 5 cases have a gravitino with mass less than 3
TeV. Thus there is most likely a gravitino problem.   In addition
the lightest moduli mass is of order (Table \ref{tab:weak_obs})
several 100s GeV.  Thus there is also a cosmological moduli problem.
But there is hope.  The next lightest massive modulus [nLMM] has, in
all cases, a mass above 20 TeV.   A detailed cosmological analysis
is beyond the scope of this paper.  However, it is possible that
when cosmological temperatures are of order $m_{\text{nLMM}}$, the
universe becomes nLMM dominated.   By the time the nLMM decays all
matter is diluted and then the universe reheats to temperatures
above the scale of big bang nucleosynthesis (for example, see
\cite{Acharya:2008bk}).   Thus it is possible that the nLMM solves
both the gravitino and light moduli problems. Of course, then the
issue of obtaining the correct baryon asymmetry of the universe and
the dark matter abundance must be addressed. Both can in principle
be obtained via non-thermal processes at low temperature.

In Table \ref{tab:param_space} we analyze the dependence of our
results on the value of $\tan\beta$ and $\text{sgn}(\mu)$ with all
other input parameters fixed.   We find that only the value of the
light Higgs mass is sensitive to varying $\tan\beta$.  Note the
lowest value of $\tan\beta$ is obtained by the Higgs mass bound,
while the largest value of the light Higgs mass is obtained with the
largest value of $\tan\beta$ (for both signs of $\mu$).
Additionally, at large $\tan\beta$ for $\mu < 0$ the Higgs potential
becomes unbounded from below.  For $\mu > 0$ we limited the analysis
to $\tan\beta \leq 50$. The light Higgs mass does not go above 122
GeV for $\tan\beta \leq 50$.

%================================================================
\begin{table}[t!]
\renewcommand\arraystretch{1.2}
\begin{center}
\noindent\makebox[\textwidth]{ \scalebox{0.75}{
\begin{tabular}{|cc|c|c|c|c|c|c|c|c|c|c|}
\hline
\multicolumn{2}{|c|}{\multirow{2}{*}{Case}}&\multicolumn{2}{c|}{1}&\multicolumn{2}{c|}{2}&\multicolumn{2}{c|}{3}&\multicolumn{2}{c|}{4}&\multicolumn{2}{c|}{5} \\
\cline{3-12}
&&$\mu < 0$&$\mu > 0$&$\mu < 0$&$\mu > 0$&$\mu < 0$&$\mu > 0$&$\mu < 0$&$\mu > 0$&$\mu < 0$&$\mu > 0$ \\
\hline
\multirow{2}{*}{$\tan\beta$} & lo &  5 &  6 &  8 & 12 &  9 & 11 &  5 &  4 &  5 &  3 \\
                             & hi & 38 & 50 & 36 & 50 & 39 & 50 & 32 & 48 & 39 & 50 \\
\hline
\multirow{2}{*}{$m_{h^0}$} & lo & 113.5 & 113.2 & 112.4 & 112.4 & 112.4 & 112.4 & 112.4 & 116.2 & 113.5 & 112.5 \\
                           & hi & 117.4 & 117.2 & 113.7 & 113.4 & 113.6 & 113.7 & 120.5 & 121.8 & 120.8 & 121.9 \\
\hline
Neut. &   & bino & bino & bino & bino & bino & bino & bino & bino & bino & bino \\
comp. &   & $\gtrsim 99\%$ &  $\gtrsim 99\%$ & $\gtrsim 99\%$ & $\gtrsim 99\%$ & $\gtrsim 99\%$ & $\gtrsim 99\%$ & $\gtrsim 99\%$ & $\gtrsim 99\%$ & $\gtrsim 99\%$ & $\gtrsim 99\%$ \\
\hline
$m_{\tilde{\chi}^0_1}$  & lo & 149.1 & 148.5 & 82.5 & 82.0 & 98.6 & 98.2 & 68.9 & 67.3 & 53.6 & 51.5 \\
(GeV)                   & hi & 151.5 & 149.9 & 84.0 & 82.9 & 99.8 & 98.7 & 69.6 & 70.3 & 55.1 & 55.9 \\
\hline
$m_{\tilde{\chi}^{\pm}_1}$ & lo & 290.4 & 291.3 & 162.3 & 162.3 & 193.5 & 193.6 & 139.4 & 134.3 & 110.3 & 103.3 \\
(GeV)                      & hi & 298.8 & 293.7 & 167.1 & 163.7 & 197.7 & 194.5 & 141.0 & 141.6 & 113.8 & 114.6 \\
\hline
\hline
\end{tabular}
}} \caption{\label{tab:param_space}Scan over $\tan\beta$ and
$\text{sgn}(\mu)$.}
\end{center}
\end{table}
%================================================================

\FloatBarrier

%==============================================================
%==============================================================
\section{Conclusions}
\label{sec:conclusion}
%==============================================================
%==============================================================

As a candidate theory of all fundamental interactions, string theory
should admit at least one example of a four-dimensional vacuum which
contains particle physics and early universe cosmology consistent
with the two standard models.  In this context, the recently found
``mini-landscape" of heterotic orbifold
constructions~\cite{Buchmuller:2005jr, Buchmuller:2006ik,
Lebedev:2006kn, Lebedev:2007hv, Lebedev:2008un} provide us with very
promising four-dimensional perturbative heterotic string vacua.
Their low-energy effective field theory was shown to resemble that
of the MSSM, assuming non-zero VEVs for certain blow-up moduli
fields which parametrize resolutions of the orbifold fixed points
along $F$- and $D$-flat directions in global supersymmetry.

In this paper we have dealt with the task of embedding the globally
supersymmetric constructions of the heterotic ``mini-landscape" into
supergravity and then stabilizing the moduli of these
compactifications, including their orbifold fixed point blow-up
moduli. The blow-up moduli appear as chiral superfields contained in
the twisted sectors of the orbifolded heterotic string theory.  They
are singlets under all standard model gauge groups, but are charged
under several unwanted $U(1)$ gauge symmetries, including the
universal anomalous $U(1)_A$ gauge symmetry of the heterotic string.
Note, moduli stabilization of string compactifications is a crucial
precondition for comparing to low energy data, as well as for
analyzing any early universe cosmology, such as inflation, in a
given construction.

Section \ref{sec:general_structure} served the purpose of reviewing
the ingredients and structure of the heterotic 4d $N=1$
supergravity inherited from orbifold compactifications of the 10d
perturbative $E_8\otimes E_8$ heterotic string theory. The general
structure of these compactifications results in:
\begin{itemize}
\item[i)] a standard no-scale K\"{a}hler potential for the bulk volume and
complex structure moduli, as well as the dilaton, together with
\item[ii)] gaugino condensation in the unbroken sub-group of the hidden $E_8$, and
\item[iii)] the fact that the non-perturbative (in the world-sheet instanton sense)
Yukawa couplings among the twisted sector singlet fields contain
terms explicitly breaking the low-energy $U(1)_R$-symmetry.
\end{itemize}

We have shown in Section \ref{sec:one_gaugino_condensate} that these
three general ingredients, present in all of the ``mini-landscape"
constructions, effectively realize a KKLT-like setup for moduli
stabilization. Here, the existence of terms explicitly breaking the
low-energy $U(1)_R$-symmetry at high order in the twisted sector
singlet fields is the source of the effective small term $w_0$ in
the superpotential, which behaves like a constant with respect to
the heterotic dilaton \cite{Kappl:2008ie}. Utilizing this, the
presence of just a single condensing gauge group in the hidden
sector (in contrast to the racetrack setups in the heterotic
literature) suffices to stabilize the bulk volume $T$ (and, by
extension, also the bulk complex structure moduli $U$), as well as
the dilaton $S$ at values $\langle {\rm Re}\;T\rangle \sim 1.1 -
1.6$ and $\langle{\rm Re}\;S\rangle\sim 2$. These are the values
suitable for perturbative gauge coupling unification into $SU(5)$-
and $SO(10)$-type GUTs distributed among the orbifold fixed points.
Note, we have shown this explicitly for the case one $T$ modulus and
a dilaton, however, we believe that all bulk moduli will be
stabilized near their self-dual points
\cite{Font:1990nt,Cvetic:1991qm}.

At the same time, the near-cancelation of the $D$-term of the
universal  anomalous $U(1)_A$-symmetry stabilizes non-zero VEVs for
certain gauge invariant combinations of twisted sector singlet
fields charged under the $U(1)_A$. This feature in turn drives
non-vanishing $F$-terms for some of the twisted sector singlet
fields.  Thus, together with the $F$-terms of the bulk volume moduli
inherited from modular invariance, it is sufficient to uplift the
AdS vacuum to near-vanishing cosmological constant.

The structure of the superpotential discussed in this paper, $\fw
\sim w_0 e^{- b T} + \phi_2 \ e^{- a S - b_2T}$, behaves like a
`hybrid KKLT' with a single-condensate for the dilaton S, but as a
racetrack for the $T$ and, by extension, also for $U$ moduli.  An
additional matter $F_{\phi_2}$ term driven by the cancelation of the
anomalous $U(1)_A$ $D$-term seeds successful up-lifting.

We note the fact that the effective constant term in the
superpotential, $w_0$, does not arise from a flux superpotential akin to the
type IIB case.  This leaves open (for the time being) the question
of how to eventually fine-tune the vacuum energy to the
$10^{-120}$-cancelation necessary.

Section \ref{sec:moduli_stabilization} then serves to demonstrate
how the success of stabilizing the bulk moduli and breaking
supersymmetry in the $F$-term sector, driven by the $U(1)_A$ $D$-term
cancelation, transmits itself to the chiral singlet fields from the
untwisted and twisted sectors of the orbifold compactification which
contain, among others, the blow-up moduli associated with the
orbifold fixed points. The effects from the bulk moduli
stabilization and supersymmetry breaking, transmitted through
supergravity, generically suffice to stabilize all of the twisted
sector singlet fields at non-zero VEVs.   This property was assumed
in the original ``mini-landscape" construction in order to decouple
the non-MSSM vector-like exotic matter, and our arguments provide the
first step towards a self-consistent justification for these
assumptions.

In Section \ref{sec:spectrum} we estimate the structure of the soft
terms from the moduli sector supersymmetry breaking at the high
scale. We find that the contributions from high-scale gauge
mediation are subdominant (although not parametrically suppressed)
compared to the gravity mediated contributions. Upon RGE running the
high-scale soft terms to the weak scale using {\tt softSUSY}, we
obtain several benchmark patterns of sparticle and Higgs masses (see
Table \ref{tab:weak_obs}). The low-energy spectrum features an
allowed window of $\tan\beta$ values for $m_{3/2}<5\,{\rm TeV}$. It
generically contains a light chargino/ neutralino spectrum and heavy
squarks and sleptons. The lightest MSSM partner, in the 5 benchmark
cases studied, is given by a bino ($> 99\%$) with mass $\gtrsim 52
\,{\rm GeV}$.  If this were the LSP, it would yield a dark matter
abundance which over closes the universe, however, the
``mini-landscape'' models offer some possible resolutions.  One
possibility is that the bino decays into an axino, the partner of
the invisible axion responsible for canceling the $\theta$-angle of
QCD, which is present in many of the ``mini-landscape" setups
\cite{Choi:2009jt}. We have also considered an alternative
possibility that the late decay of the next to lightest massive
modulus might ameliorate or solve the cosmological gravitino and
moduli problem.  This would then dilute the above mentioned
cosmological abundance of binos.  Of course, the non-thermal
production of dark matter and a baryon asymmetry must then be
addressed.  Note, however, the resolution of these cosmological
questions are beyond the scope of the present paper.

Summarizing, we have given a mechanism for moduli stabilization and
supersymmetry breaking for the perturbative heterotic orbifold
compactifications.  It relies on the same variety and number of
effective ingredients as the KKLT construction of type IIB flux
vacua and thus represents a significant reduction in necessary
complexity, compared to the multi-condensate racetrack setups
utilized so far.  When applied to a simplified analog of the
``mini-landscape" heterotic orbifold compactifications, which give
the MSSM at low energies, it leads to fully stabilized 4d heterotic
vacua with broken supersymmetry and a small positive cosmological
constant.  Moreover, most of the low energy spectrum could be
visible at the LHC.

We leave some important questions like the problem of the full
fine-tuning of the vacuum energy to near-vanishing, or the existence
of an inflationary cosmology within these stabilized
``mini-landscape" constructions for future work. Further study is also warranted with respect to potential cosmological moduli and gravitino problems that may be associated with sub-100 TeV moduli and gravitino mass values (see e.g.~\cite{gravitinoprob}).  Finally,  the
numerical evaluation of any particular ``mini-landscape" vacuum
requires analyzing the supergravity limit with three bulk moduli,
$T$, one bulk complex structure modulus, $U$, and of order 50
blow-up moduli. A detailed analysis of this more realistic situation
would require a much better handle on the moduli space of heterotic
orbifold models than is presently available.

\newpage

%==============================================================
%==============================================================
\section*{Acknowledgements}
%==============================================================
%==============================================================

A.W. would like to thank the Physics Department at the Ohio State
University for their warm hospitality, where part of this work was
completed. B.D. and S.R. would like to thank the Stanford Institute
for Theoretical Physics for their hospitality, where the bulk of
this work was completed.  Also  B.D., S.R. and A.W. thank the KITP
at UC Santa Barbara for support.  We are also indebted to Patrick
Vaudrevange for aid with the superpotential for Model IA. B.D. and
S.R. are supported by DOE grant DOE/ER/01545-884. A.W. is supported
in part by the Alexander-von-Humboldt foundation, as well as by NSF
grant PHY-0244728. This research was supported in part by the National Science Foundation under Grant No.~NSF PHY05-51164. We would also like to thank K.~Bobkov and
N.~Craig for useful discussions.  S.R. would also like to thank
W.~Buchm\"{u}ller, J.~Conlon, E.~Dudas, G.~Dvali, T.~Kobayashi, D.~L\"{u}st, H.P.~Nilles, M.~Ratz, S.~Ramos-Sanchez, and G.G.~Ross for discussions.

%==========================================================
%----------------------------------------------------------
%==========================================================
\clearpage\newpage
\appendix
%==========================================================
%----------------------------------------------------------
%==========================================================

%==============================================================
%==============================================================
\section{A Different Racetrack.} \label{app:diff_racetrack}
%==============================================================
%==============================================================

The form of the gaugino condensate, given in Eqn.
(\ref{duality_invariant_gaugino_condensate}), ensures that the
non-perturbative part of the superpotential is invariant under the
modular group \slz.  In deriving the form of $\fw_{\textsc{np}}$,
however, we have neglected the fact that the presence of discrete
Wilson lines often break the modular group \slz to one of its
subgroups.  It has been noted \cite{Love:1996sk} that turning on one
or more Wilson lines breaks the modular group $\slz$ down to one of
its subgroups.  Define the subgroup $\Gamma_0(p) \subset \slz$.  The
subgroup is defined as the set of $2\times 2$ matrices such
that\footnote{A detailed mathematical treatment of the modular
functions can be found in Reference \cite{modular}.}
\begin{eqnarray}
   \mathcal{M} &\equiv& \left(\begin{array}{cc}a&b\\c&d\end{array}\right), \\
   ad-cb &=& 1, \\
    a,b,c,d&\in&\mathbb{Z}, \\
    c &\equiv& 0 \mod p, \,\,\, p\in\mathbb{P},
\end{eqnarray}
where $\mathbb{P}$ is the set of prime integers.  Under this
subgroup, then, the invariant function is a linear combination of
Dedekind $\eta$ functions:
\begin{equation}
   f_{p}(\tau) = \frac{1}{p}\sum_{\lambda=0}^{p-1} \eta\left(\frac{\tau+\lambda}{p}\right).
\end{equation}

%==============================================================
\section{The Role of Holomorphic Monomials} \label{app:HIM}
%==============================================================

Supersymmetry can be broken by either $F$ terms or $D$ terms.  In a
generic supersymmetric gauge theory, $D=0$ is satisfied only along
special directions in moduli space.  These directions are described
by holomorphic, gauge invariant monomials (HIMs) \cite{Luty:1995sd,
Cleaver:1997jb, Dundee:2008gr}.  The moduli space of a general
heterotic string model is significantly more complex than that of
our simple models.  Not only are there many more fields in the
picture, there are also many more gauge groups.

Consider a theory with gauge symmetry $\U1^{\rho} \otimes \U1_A$,
where $A$ stands for anomalous.  The $D=0$ constraints are
\begin{equation} \label{generic_D_term}
   D_{a\neq A} \sim \sum_{i} q_i^a \left|\phi_i\right|^2 = 0.
\end{equation}
A generic HIM can be written in terms of fields $\phi_i$ with
charges $q_i^j$
\begin{equation} \label{general_HIM}
   \mathcal{H}\left[\phi_i\right] = \prod_{i} \phi_i^{n_i}, \,\,\, n_i > 0,
\end{equation}
such that
\begin{equation}
   \sum_i n_i q_i^j = 0, \,\,\, \forall j \neq A.
\end{equation}
The requirement that $n_i > 0$ is a reflection of the holomorphicity
of $\mathcal{H}$, while the requirement that the sum over $n_i$
(weighted by the charges) vanishes is a reflection of the gauge
invariance.  The general HIM in Eqn. (\ref{general_HIM}) relates the
VEVs of the fields $\phi$ as follows:
\begin{equation} \label{HIM_VEV_relation}
   \frac{\left|\phi_1\right|}{\sqrt{n_1}} = \frac{\left|\phi_2\right|}{\sqrt{n_2}} = \cdots.
\end{equation}
Given this relationship, one can show that the Eqns.
(\ref{generic_D_term}) can be satisfied.  Notice that no scale is
introduced in Eqn. (\ref{HIM_VEV_relation}): the HIMs (in general)
only constrain the relative magnitudes of the $\phi$ VEVs, and gives
no information about their phases or their absolute magnitudes.

The procedure for dealing with an anomalous $\U1_A$ works the same
way.  Instead of Eqn. (\ref{generic_D_term}), one has
\begin{equation}
    D_{A} \sim \sum_{i} q_i^A \left|\phi_i\right|^2 + \xi = 0,
\end{equation}
and we will assume that $\xi > 0$.  In this case, one needs to find
a monomial which is holomorphic and gauge invariant under all of the
$\rho$ \U1 factors, but which carries a net negative charge under
the anomalous $\U1_A$ \cite{Luty:1995sd, Cleaver:1997jb,
Dundee:2008gr}.  The situation is different than the case with
non-anomalous symmetries, as a mass scale is introduced into the
problem.

In a heterotic string orbifold, the FI term is generated by the
mixed gauge-gravitational anomaly, and is canceled by the
Green-Schwarz mechanism, which forces singlets to get VEVs of order
the FI scale (typically $\sim \mstring$).  Usually, several singlets
participate in this cancellation, all receiving VEVs of the same
order.   In the ``mini-landscape" models~\cite{Lebedev:2007hv},
supersymmetric vacua were obtained, prior to the consideration of
any non-perturbative effects.   A holomorphic gauge invariant
monomial was found which is invariant under all other $U(1)$s but
with net charge under $U(1)_A$ opposite to that of the FI term. This
composite field necessarily gets a non-zero VEV to cancel the FI
term.   Our field $\phi_2$ in the simple model gives mass to the
vector-like exotics of the hidden sector and thus it also appears in
the non-perturbative superpotential.   In a more general heterotic
model, $\phi_2$ would be replaced by an HIM which also cancels the
FI term.

\bibliographystyle{utphys}
\bibliography{susybreaking_v2}

\end{document}